\documentclass[a4paper,fleqn,usenatbib]{mnras}

\usepackage{mathptmx}

\usepackage[T1]{fontenc}
\usepackage{ae,aecompl}

\usepackage{graphicx}	
\usepackage{amsmath}	
\usepackage{amssymb}	
\usepackage{pdflscape}  
\usepackage{bm}

\title[Primordial gravitational waves with CMB B-modes]{Optimistic estimation on probing primordial gravitational waves with CMB B-mode polarization}

\author[Q.-G. Huang and S. Wang]{
Qing-Guo Huang$^{1,2,3,4}$\thanks{e-mail: huangqg@itp.ac.cn},
        Sai Wang$^{5}$\thanks{e-mail: wangsai@itp.ac.cn.}\\
$^{1}$CAS Key Laboratory of Theoretical Physics, Institute of Theoretical Physics, Chinese Academy of Sciences, Beijing 100190, China\\
$^{2}$School of Physical Sciences, University of Chinese Academy of Sciences, No. 19A Yuquan Road, Beijing 100049, China\\
$^{3}$Center for Gravitation and Cosmology, College of Physical Science and Technology, Yangzhou University, Yangzhou 225009, China\\
$^4$Synergetic Innovation Center for Quantum Effects and Applications, Hunan Normal University, Changsha 410081, China\\
$^{5}$Department of Physics, The Chinese University of Hong Kong, Shatin, N.T., Hong Kong 999077, China}

\date{Accepted XXX. Received YYY; in original form ZZZ}

\pubyear{2018}

\begin{document}
\label{firstpage}
\pagerange{\pageref{firstpage}--\pageref{lastpage}}
\maketitle

\begin{abstract}
In the measurements of cosmic microwave background polarizations, three frequency channels are necessary for discriminating the primordial B-modes from the polarized dust and the synchrotron emission. We carry out an optimistic estimate on the sensitivity to the detection of primordial gravitational waves using the cosmic microwave background B-modes only, and explore how to reach the thresholds for the tensor-to-scalar ratio $r$ in the theoretically well-motivated inflation models. For example, Lyth bound implies $r \simeq 2\times10^{-3}$, a corrected Lyth bound shows $r \simeq 7\times10^{-4}$, and some typical inflation models gives $r\simeq4\times10^{-5}$. Taking into account the up-to-date constraints on $r$, i.e. $r_{0.05}<0.07$ at $95\%$ confidence, we find that the consistency relation $n_t=-r/8$ in the canonical single-field slow-roll inflation cannot be distinguished from the scale invariance, due to the cosmic variance in the power spectrum of cosmic microwave background B-modes. The cosmic variance places an inevitable limit on the measurements of the tensor spectral index, i.e. $\sigma_{n_t}\simeq0.01$ for $2\leqslant\ell\leqslant \ell_\text{max}=300$. 
\end{abstract}




\section{Introduction}
\noindent
The inflation model \citep{Starobinsky:1979ty,Starobinsky:1980te,Guth:1980zm,Linde:1981mu,Albrecht:1982wi,Sato:1980yn} has been the leading paradigm of the very early Universe in past three decades. Not only does it resolve the flatness, horizon and monopole problems in the hot big-bang theory, but also seeds primordial cosmological perturbations which evolved into the large-scale structures in the Universe \citep{Mukhanov:1990me}. The primordial tensor perturbations predicted by the inflation are called as the primordial gravitational waves. In general, the power spectrum of primordial gravitational waves is parameterized as a power-law form, namely,
\begin{equation}\label{pgw}
P_t(k)=rA_s\left(\frac{k}{k_{{p}}}\right)^{n_t}
\end{equation}
where $r$ is called the tensor-to-scalar ratio at the pivot scale $k_p$, $n_t$ denotes the tensor spectral index or the tensor tilt, $A_s$ is the amplitude of power spectrum of primordial scalar perturbations.
This is a dimensionless power spectrum, which is also denoted by $\Delta_{t}^{2}$ in the literature.
The tensor tilt characterizes the scale dependence of the tensor power spectrum.
In the simplest class of inflation, i.e. the canonical single-field slow-roll models, there is a consistency relation between $r$ and $n_t$, i.e. $n_t=-r/8$ \citep{Liddle:1992wi}. Therefore, the tensor power spectrum is red-tilted.
A more generic formula in the single-field inflation can be found in Reference~\citep{Kobayashi:2011nu}.
In the effective field theory of single-field inflation, at leading order in derivatives, the tensor power spectrum is fixed by the Hubble rate during inflation \citep{Creminelli:2014wna}. Therefore, detecting a positive tilt would imply $\dot{H}>0$ and a violation of the null energy condition, which is difficult to realize without incurring in instabilities.
However, a positive tensor tilt would be allowed if some higher-derivative operators are included \citep{Baumann:2015xxa}.

Primordial gravitational waves leave several characteristic fingerprints on the CMB, including the temperature anisotropies and the E/B-mode polarizations \citep{Grishchuk:1974ny,Starobinsky:1979ty,Rubakov:1982df,Crittenden:1993ni,Kamionkowski:1996zd,Kamionkowski:1996ks,Zaldarriaga:1996xe,Hu:1997mn}. Using the CMB temperature anisotropies only, \emph{Planck} 2015 results (P15) \citep{Ade:2015lrj} showed an indirect upper limit on the tensor-to-scalar ratio, namely, $r_{0.05}<0.12$ at 95\% confidence level (CL). Using the B-mode data, BICEP2 \& Keck Array (BK14) \citep{Array:2015xqh} gave the latest direct upper bound $r_{0.05}<0.09$ at 95\% CL. Combining the above two datasets with other low-redshift datasets, the upper bound becomes tighter, i.e., $r_{0.05}<0.07$ at 95\% CL \citep{Array:2015xqh,Huang:2015cke} \footnote{In 2018, BICEP \& Keck Array \citep{Ade:2018gkx} released results from an analysis of all data up to and including 2015 observing season. The data analysis yielded a constraint $r_{0.05}<0.07$ at 95\% CL, which tightened to $r_{0.05}<0.06$ by combining with Planck temperature measurements and other datasets. These are the strongest constraints to date on the tensor-to-scalar ratio $r$.}. Using the gravitational-wave observations only, a recent constraint on the tensor tilt was found to be $n_t=-0.76^{+1.37}_{-0.52}$ at $68\%$ CL \citep{Huang:2015gka}. Further combining with the indirect observations of gravitational waves, this constraint becomes $n_t=-0.05^{+0.58}_{-0.87}$ at $95\%$ CL \citep{Cabass:2015jwe}. Both constraints on the tensor tilt are consistent with the scale-invariant spectrum. Reference \citep{Meerburg:2015zua} obtained similar results.

\begin{figure}
\centering
\includegraphics[width=\columnwidth]{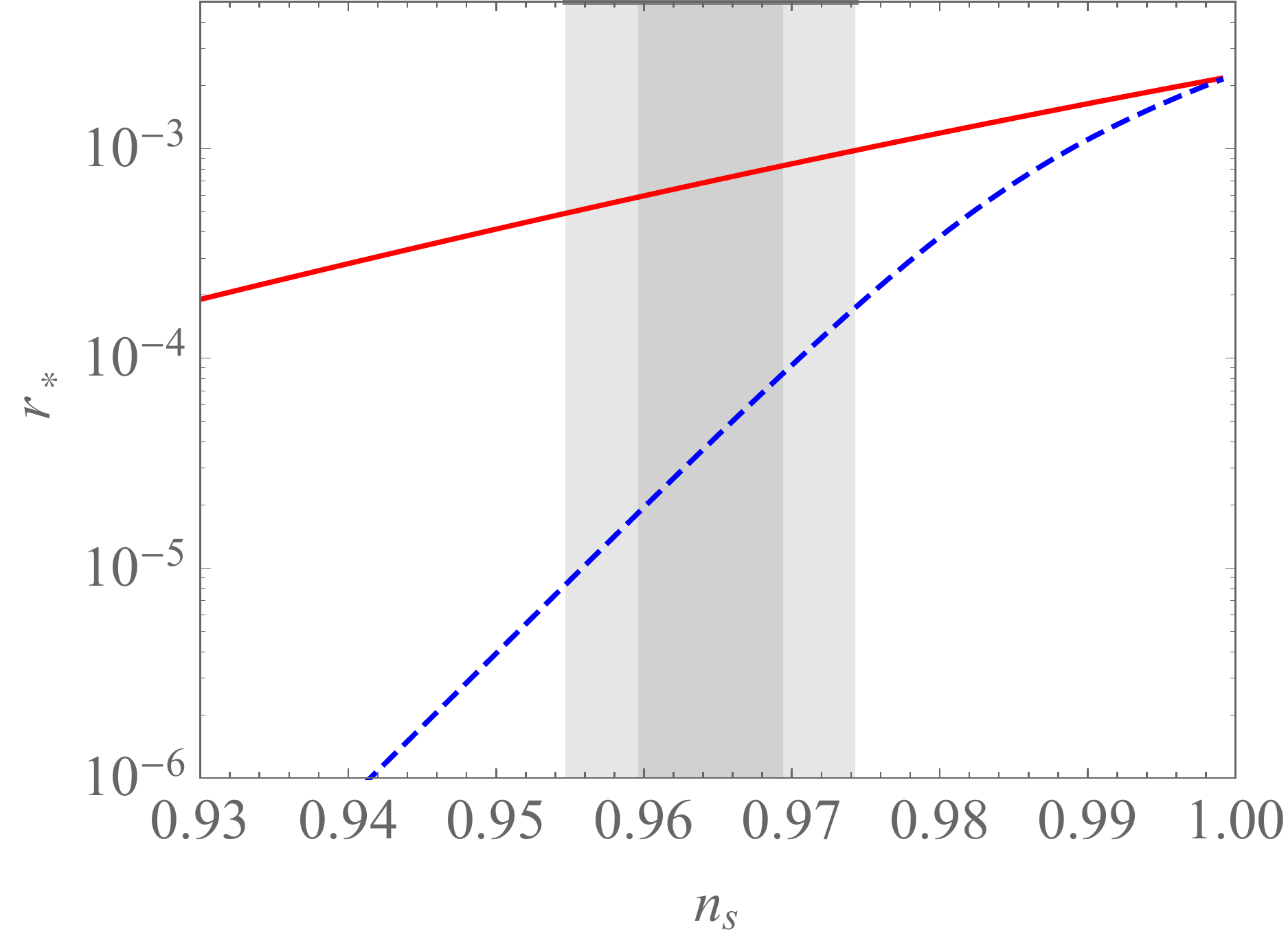}
\caption{\small The threshold for the tensor-to-scalar ratio corresponding to $|\Delta\phi|/M_p=1$. The red solid curve and blue dashed curve, respectively, correspond to the general single-field slow-roll inflation model and some typical inflation model in which $r(N)=16\alpha/(N+\alpha^{1/p})^p$. The gray and light gray bands, respectively, correspond to the  $68\%$ and $95\%$ limits on $n_s$ from \emph{Planck} data released in 2015 \citep{Ade:2015lrj}. }
\label{fig:lythb}
\end{figure}

Since the primordial gravitational waves have not been detected, it is worthy to explore some well-motivated thresholds for the tensor-to-scalar ratio $r$. In Reference \citep{Lyth:1996im}, Lyth found that the tensor-to-scalar ratio is related to the excursion distance of inflaton during the inflationary era through
\begin{equation}
{|\Delta\phi|/ M_p}=\int_0^{N_*} \sqrt{r(N)/ 8} ~dN\ ,
\end{equation}
where $M_p=1/\sqrt{8\pi G}$ is the reduced Planck scale and $N$ is the e-folding number before the end of inflation. If $r$ is a constant, the threshold $|\Delta\phi|/M_p=1$ implies $r=8/N_*^2\simeq 2\times 10^{-3}$.
However, the current constraint on $n_s$ from P15 \citep{Ade:2015lrj} showed $n_s=0.9645\pm 0.0049$ at $68\%$ CL, and the tensor-to-scalar ratio $r$ evolves with the expansion of the Universe during the inflationary era, namely $-{d\ln r/ dN}=n_t-(n_s-1)$ by definition. For $n_s\neq 1$, the threshold for $r$ is modified to be $r_*\simeq 2(1-n_s)^2/[\exp((1-n_s)N_*/2)-1]^2$ \citep{Huang:2015xda}, i.e. the red solid curve in Fig.~\ref{fig:lythb}, and then $r_*\simeq 7\times 10^{-4}$ for $n_s=0.9645$. Furthermore, the tensor-to-scalar ratio takes the form of $r(N)=16\alpha/(N+\alpha^{1/p})^p$ in some typical inflation models \citep{Huang:2015cke,Huang:2007qz,Garcia-Bellido:2014wfa}, and the threshold for $r$ in this case is the blue dashed curve in Fig.~\ref{fig:lythb}. 
For $n_s=0.9645$, we have $r_*\simeq 4\times 10^{-5}$.
In addition, from Fig.~\ref{fig:lythb}, the thresholds for $r$ in both cases approach to $2\times 10^{-3}$ when $n_s\rightarrow 1$.

Two questions we are facing are how to reach these well-motivated theoretical thresholds for the tensor-to-scalar ratio and how we can test the consistency relation $n_t=-r/8$ for the simplest class of inflation model.
Future improvements to the sensitivity on the primordial gravitational waves should mainly come from the CMB polarization experiments. In the realistic observations, however, the foreground radiations significantly contaminate the primordial B-mode polarizations \citep{Adam:2014bub,Mortonson:2014bja,Flauger:2014qra,Cheng:2014pxa,Ade:2015tva,Aghanim:2016cps}. They should be taken into consideration seriously. A recent study \citep{Creminelli:2015oda} showed that the theoretically motivated $r\sim2\times10^{-3}$ can be achieved by some future experiments if the instrumental white noise is reduced to $\sim1\mu$K-arcmin and the lensing B-modes reduced to $10\%$. Their forecasts are not changed significantly with respect to the previous estimates \citep{Lee:2014cya}. However, Reference  \citep{Huang:2015gca} shows that these experiments are not sensitive enough to discriminate the consistency relation in the canonical single-field slow-roll inflation models from the scale-invariant spectrum. 
Other related studies can be found, for example, in References~\citep{Cabass:2015jwe,Escudero:2015wba,Errard:2015cxa,Kamionkowski:2015yta,Zhao:2015sla,Guzzetti:2016mkm,Lasky:2015lej,Wang:2016tbj} and references therein.

In this paper, we will explore several potential setups for the CMB polarization experiments, for which we generally consider the contamination including foreground emissions, white noise, and CMB lensing, regardless of the specific methods to reduce them. We will take these contamination into consideration to estimate the projected sensitivity of these experiments to the detection of the primordial gravitational waves. In particular, we will explore how to reach the theoretically motivated thresholds for the tensor-to-scalar ratio, and check the possibility to test the consistency relation in the simplest class of inflation. 



The rest of the paper is arranged as follows. In section \ref{sec:methodologhy}, we describe what is the methodology to deal with primordial B-modes, foregrounds, instrumental noise and CMB lensing. The Fisher information matrix is also briefly introduced. In section \ref{sec:analysisresults}, we show the results of our analysis. The conclusion and discussion are listed in section \ref{sec:conclusion}.

\section{Methodology}
\label{sec:methodologhy}
\noindent
In this paper, we only focus on the CMB B-mode polarization, regardless of the temperature anisotropies and the E-mode polarization. In this section, we demonstrate what are the methods to deal with various components of the CMB B-modes and the technique of Fisher information matrix.

\subsection{Primordial B-modes}
\noindent
Using the publicly available \texttt{CAMB} program package \citep{Lewis:1999bs,Howlett:2012mh}, we can numerically calculate the angular power spectrum of the primordial B-modes in the linear perturbation theory, given a spectrum of primordial gravitational waves. The result is 
\begin{equation}
C_\ell^{BB}=\int d\ln k~P_t(k)~[\Delta^{B}_{\ell}(k)]^{2}\ ,
\end{equation}
where $\Delta^{B}_{\ell}$ is the transfer function and $P_t(k)$ denotes the power spectrum of primordial gravitational waves. The formula of $P_t(k)$ is given by Eq.~(\ref{pgw}). In the following, we use $\tilde{C}_{\ell}={\ell(\ell+1)}C^{BB}_{\ell}/{(2\pi)}$ instead of $C^{BB}_{\ell}$ for convenience.

Based on the \emph{Planck} 2015 results \citep{Adam:2016hgk,Ade:2015xua}, the six independent cosmological parameters in the base $\Lambda$CDM fiducial model are fixed to their best-fit values at the scalar pivot scale $k_p=0.05~\textrm{Mpc}^{-1}$, namely,
\([\Omega_b h^2,\Omega_c h^2,100\theta_{MC},\tau,\ln(10^{10}A_s),n_s]=[0.02225,0.1198,1.04077,0.058,3.094,0.9645]\).
They include the baryon density today $(\Omega_b h^2)$, the cold dark matter density today $(\Omega_c h^2)$, the angular scale of the sound horizon at last-scattering ($\theta_{\rm MC}$), the Thomson scattering optical depth due to the reionization $(\tau)$, the amplitude of scalar power spectrum $(A_s)$, and the spectral index of scalar power spectrum $(n_s)$. 
For the canonical single-field slow-roll inflation, the consistency relation is  $n_t=-r/8$ \citep{Liddle:1992wi}. 
Considering the current upper bounds on $r$, the absolute value of the tensor tilt is expected to be less than $\mathcal{O}(10^{-2})$. 
For simplicity, we can thus set a vanishing fiducial value to $n_t$. 
For consistency, however, we still fix the tensor tilt to the consistency relation in this work.


\subsection{Foregrounds}
\noindent
For the foregrounds, we consider the synchrotron (S) and the Galactic polarized dust (D). One can separate them from each other in the power spectrum of CMB B-mode polarization, since they occupy totally different frequency dependence. At a given frequency $\nu$, the power spectra of them are usually given by
\begin{eqnarray}
S_{\ell\nu}&=&\left(W^{S}_{\nu}\right)^{2}C^{S}_{\ell}=\left(W^{S}_{\nu}\right)^{2}A_{S}\left(\frac{\ell}{\ell_{S}}\right)^{\alpha_{S}}\ ,\\
D_{\ell\nu}&=&\left(W^{D}_{\nu}\right)^{2}C^{D}_{\ell}=\left(W^{D}_{\nu}\right)^{2}A_{D}\left(\frac{\ell}{\ell_{D}}\right)^{\alpha_{D}}\ \ ,
\end{eqnarray}
where the frequency dependence are denoted by $W^{S}_{\nu}$ and $W^{D}_{\nu}$, respectively. One has the following formulae, namely,
\begin{eqnarray}
W^{S}_{\nu}&=&\frac{W^{CMB}_{\nu_{S}}}{W^{CMB}_{\nu}}\left(\frac{\nu}{\nu_{S}}\right)^{\beta_{S}}\ ,\\
W^{D}_{\nu}&=&\frac{W^{CMB}_{\nu_{D}}}{W^{CMB}_{\nu}}\left(\frac{\nu}{\nu_{D}}\right)^{1+\beta_{D}}\frac{e^{h\nu_{D}/k_{B}T}-1}{e^{h\nu/k_{B}T}-1}\ ,
\end{eqnarray}
where the foregrounds have been normalized by the CMB blackbody, i.e. $W^{CMB}_{\nu}={x^2 e^x}/{\left(e^x - 1\right)^2}$ in which we use $x=\frac{h\nu}{k_B T_{CMB}}$, and $T_{CMB}=2.7255\textrm{K}$ denotes a mean temperature of the CMB. The fiducial values of foreground parameters are listed in Tab.~\ref{tab:foregroundparameters}.
\begin{table}
\centering
\renewcommand{\arraystretch}{1.5}
\begin{tabular}{|ccc|}
 \hline
  Parameters & Synchrotron & Polarized Dust \\
\hline
$A_{72\%}[\mu K^2]$& $2.1\times10^{-5}$ &  $0.169$ \\
$A_{1\%}[\mu K^2]$& $4.2\times10^{-6}$ &  $0.006$ \\
$\nu [GHz]$& $65$ &  $353$ \\
$\ell$& $80$ &  $80$ \\
$\alpha$& $-2.6$ &  $-2.42$ \\
$\beta$& $-2.9$ &  $1.59$ \\
$T[K]$& $-$ &  $19.6$ \\
\hline
\end{tabular}
\caption{A list of foreground parameters.}
\label{tab:foregroundparameters}
\end{table}
Here $A_{f_{\rm{sky}}}$ denotes the cleanest effective area, which occupies a fraction ($f_{\rm{sky}}$) of the sky, and its unit is $\mu$K$^2$.

The synchrotron foreground was measured by WMAP satellite \citep{Page:2006hz}, and the Galactic polarized dust was measured by \emph{Planck} satellite \citep{Adam:2014bub}. The synchrotron dominates below $90~\textrm{GHz}$, while the Galactic polarized dust becomes dominant above this frequency. The foregrounds could also be cross-correlated with each other \citep{Flauger:2014qra}. To achieve a rough estimation, we assume a correlation taking the form $g\sqrt{S_{\ell \nu_i}D_{\ell \nu_j}}$, where $g$ denotes a correlation coefficient. In our fiducial model, we set $g=0.5$ which is independent of $f_{\rm{sky}}$, $\ell$ and $\nu$. More detailed analysis is beyond the scope of this paper.

\subsection{White noise}
\noindent
We consider the white noise which stems from the Fourier transformation of a Gaussian beam from the real space to the harmonic space. We do not consider the systematics depending strongly on specific experimental setups. The power spectrum of white noise is defined as \citep{Knox:1995dq}
\begin{equation}
\mathcal{N}_{\ell}=\frac{\ell(\ell+1)}{2\pi}\delta P^{2}e^{\ell^2 \sigma_b^2}\ ,
\end{equation}
where $\delta P$ denotes the instrumental sensitivity to the CMB polarizations, and $\sigma_b=\theta_{FWHM}/\sqrt{8\ln2}$ denotes the beam-size variance. In this study, we consider several experimental settings, regardless of how to specifically implement them. The sensitivity $\delta P$ is assumed to take the same magnitude for all frequency bands, and freely vary in the range $[10^{-2}, 10^{2}]$$\mu$K-arcmin. We leave this parameter to be free since we expect to study its effect on the detection of primordial B-modes and how to improve it. In addition, the experimental resolution $\theta_{FWHM}$ is fixed to be 5 arcmin here.

To get a rough sense to our setup, we note the following experimental specifications. LiteBIRD \footnote{litebird.org} will have a beam of $\sim 30$ arcmin, COrE-M5 \footnote{www.core-mission.org} will have $\sim3.7$ arcmin, and CMB-S4 \footnote{https://cmb-s4.org} will have $\sim 3$ arcmin. LiteBIRD will reach a sensitivity of $\sim3.2$$\mu$K-arcmin, COrE-M5 will reach $\sim2$$\mu$K-arcmin, and CMB-S4 is expected to reach $\sim 1$$\mu$K-arcmin. Therefore, our lower value for the power noise seems very futuristic, while the upper bound seems very pessimistic.

\subsection{Lensing residual}
\noindent
The gravitational weak lensing of the CMB provides an another source of contamination to the detection of primordial B-modes \citep{Lewis:2006fu}. One cannot deal with the lensing B-modes in the similar way as the foregrounds, since the lensing B-modes have the same frequency dependence as the primordial B-modes. However, one can reconstruct the lensing potential with the CMB temperature or E-modes at small angular scales, and then subtract the lensing effects from the CMB B-modes at larger angular scales \citep{Knox:2002pe,Kesden:2002ku,Seljak:2003pn,Smith:2010gu,Ade:2015nch}. This process is called delensing.
One can define a parameter to describe the residual of lensing B-modes after delensing, i.e. $\alpha_L$, which is also independent of $f_{\rm{sky}}$, $\ell$ and $\nu$. In other words, $1-\alpha_L$ denotes the delensing efficiency. There is not any physical limit on the delensing efficiency in principle. 

In this study, we set $\alpha_L\in[10^{-2},1]$ without considering how to specifically implement the delensing. Note that the lower bound for $\alpha_{L}$ seems very optimistic, while the upper bound seems very pessimistic. Using the procedure described in Reference \citep{Errard:2015cxa}, it is possible to show that future experiments will achieve at most $\alpha_{L}=\mathcal{O}(10^{-1})$ \citep{Renzi:2018dbq}. For example, LiteBIRD is expected to reach $\alpha_{L}\simeq0.94$, COrE-M5 will reach $\alpha_{L}\simeq0.37$, and CMB-S4 will reach $\alpha_{L}\simeq0.25$.

\begin{figure}
\centering
\includegraphics[width=\columnwidth]{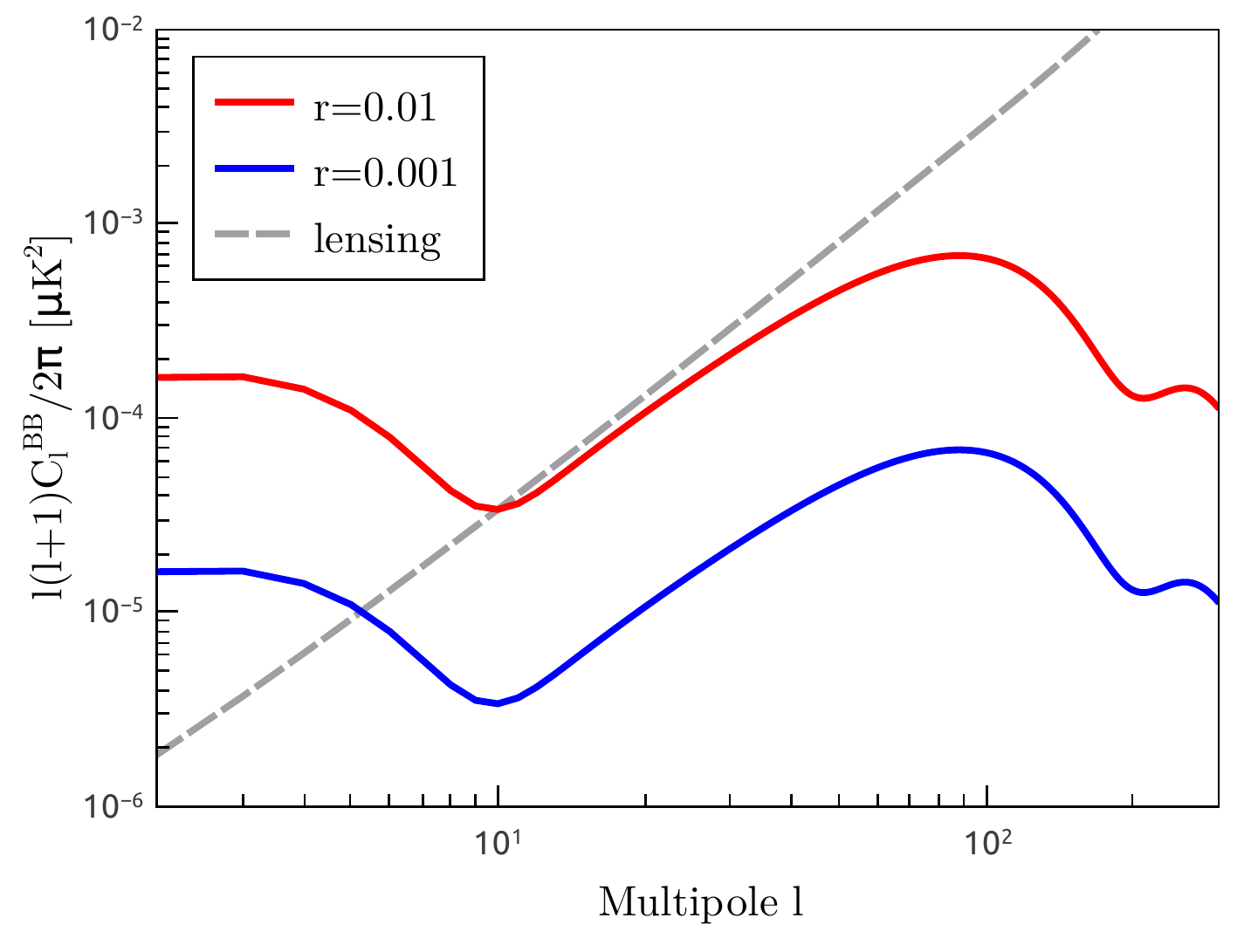}
\caption{\small The power spectra of primordial B-mode polarization (solid curves) versus the lensing B-mode spectrum (dashed curve). }
\label{fig:lensr}
\end{figure}

The lensing B-mode power can be numerically calculated using the \texttt{CAMB}. Delensing starts to become important when the tensor-to-scalar ratio is smaller than $\mathcal{O}(10^{-2})$ if we mainly concentrate on the recombination peak of B-mode power spectra, see Fig.~\ref{fig:lensr}. 
The residual power $\delta C_{\ell}^{lensing}\equiv \alpha_L C_{\ell}^{lensing}$ of the lensing B-modes could be incorporated into the power spectrum of the so-called effective noise, i.e., $\mathcal{N}_\ell\longrightarrow \mathcal{N}_\ell+\delta C_{\ell}^{lensing}$ \citep{Lee:2014cya}. In fact, the power spectrum of lensing B-modes ($\ell<150$) is similar to that of the white noise with an amplitude of $4.5\mu$K-arcmin.

\subsection{Fisher information matrix}\label{subsec:fisher}
\noindent
In general, one can use the ``component separation'' (CS) method to analyze the CMB B-mode polarization data. The average log-likelihood is given by \citep{Creminelli:2015oda}
\begin{eqnarray}\label{likelihood}
&&\hspace{-0.08\columnwidth}\langle\ln \mathcal{L}_{BB}\rangle=-\frac{1}{2}\sum_{\ell}f_{\rm{sky}}(2\ell+1)[\ln \det \left({W\mathcal{C}_{\ell}W^{T}+\mathcal{N}_{\ell}}\right)\nonumber\\
&&\hspace{0.4\columnwidth}+\textrm{tr}\left(\frac{\bar{W}\mathcal{\bar{C}}_{\ell}\bar{W}^{T}+\mathcal{N}_{\ell}}{W\mathcal{C}_{\ell}W^{T}+\mathcal{N}_{\ell}}\right)]\ ,
\end{eqnarray}
where the bar denotes the parameters fixed to their ``true'' values. Here a constant term has been discarded. $W$ describes the frequency dependence of each component of CMB B-modes. It is a $N\times 3$ matrix with a row $(1,W_{\nu_i}^D,W_{\nu_i}^{S})$. We use $N$ to denote the number of frequency bands. As a $3\times3$ matrix, $\mathcal{C}_\ell$ denotes the covariance matrix among three B-mode components. The Fisher matrix is defined as
\begin{equation}
F_{ij}=-\frac{\partial^2\langle\log\mathcal{L}_{BB}\rangle}{\partial p_i \partial p_j}\mid_{\mathbf{p}=\mathbf{\bar{p}}}\ ,
\end{equation}
where $\bar{\mathbf{p}}$ denote the ``true'' values for a set of parameters $\mathbf{p}$.
The $1\sigma$ error on a parameter $p_i$ is given by the Cramer-Rao bound, namely, $\sigma_{p_i}\geqslant\sqrt{\left(F^{-1}\right)_{ii}}$.

Taking into account the foregrounds, one of the aims of this paper is to explore how to reach the theory-motivated thresholds for the tensor-to-scalar ratio. Given $(\delta P,\alpha_L)$, we consider the likelihood as a function of six parameters $(r,A_D,A_S,\beta_D,\beta_S,g)$. We set a fiducial value $\bar{r}=0$. In this case the cosmic variance approximates zero, and the uncertainty on $r$ only depends on foregrounds, noise and lensing. Thus our estimation can be viewed as an optimistic one. Considering the uncertainties, we assume the Gaussian priors for $A_D$, $A_S$, and $\beta_S$ with the variance $50\%$, $50\%$, and $10\%$, respectively. We assume a Gaussian prior for $\beta_D$ with the variance $10\%$ for $f_{\textrm{sky}}=72\%$ while $50\%$ for $f_{\textrm{sky}}=1\%$. For other parameters, we do not assume priors any more. When forecasting $r$, we set the tensor pivot scale to be $k_p=0.01\rm{Mpc}^{-1}$ which is roughly corresponded to $\ell\simeq100$ in the CMB B-mode power spectrum. The $1\sigma$ uncertainty on $r$ is obtained by marginalizing over all other parameters.

The parameter $n_t$ will be added when one studies the consistency relation or equivalently the scale dependence of the tensor spectrum. In this case, the fiducial value of $r$ can not be vanishing any more. One should set a non-zero fiducial value for $r$. Otherwise, both $n_t$ and its $1\sigma$ uncertainty can take any real value. To be consistent with current observational limits on $r$, i.e. $r_{0.05}<0.07$ at 95\% CL, we set $\bar{r}=0.05$ and $\bar{n}_t=-\bar{r}/8$ as the fiducial value. We also study the fiducial model with $\bar{n}_{t}=0$, and the results are not changed significantly. The $1\sigma$ uncertainties on $r$ and $n_t$ are obtained by marginalizing over all other parameters. In addition, there could be some degeneracy between $r$ and $n_t$ in the $(r,n_t)$ confidence ellipse. One should choose a suitable pivot scale $k_p$ to eliminate this degeneracy \citep{Zhao:2009mj,Zhao:2011zb,Huang:2015gca}. Or equivalently, one should find a suitable pivot scale such that $(F^{-1})_{r n_t}\simeq0$. In this way, the $1\sigma$ uncertainty on $r$ will be minimal.

\section{Analysis and Results}
\label{sec:analysisresults}
\noindent
In this section, we show our results about potential uncertainties on the tensor-to-scalar ratio $r$ and the tensor tilt $n_t$ due to foregrounds, white noise and delensing residual. In particular, we predict an inevitable uncertainty on $n_t$ due to the cosmic variance of the power spectrum of CMB B-mode polarization.

We explore the CMB B-mode observations with three different frequency bands. We consider the following three combinations of frequency bands. The first case is given by the frequency bands ($35$, $90$, $350$)~GHz, the second one ($90$, $220$, $350$)~GHz and the third one ($35$, $45$, $90$)~GHz. The frequency $90$~GHz is the CMB band, $35$~GHz and $45$~GHz the synchrotron bands, and $220$~GHz and $350$~GHz the polarized dust bands. The $350$~GHz band cannot be implemented on the ground, while it can be implemented in the satellite or on a balloon. Two dust/synchrotron bands can constrain the amplitude and spectral index of the dust/synchrotron power spectrum simultaneously. In addition, we consider two different sky coverage $f_{\textrm{sky}}$ when we estimate the uncertainty on $r$. One is $72\%$, and the other one is $1\%$. We consider the CMB B-mode power spectrum with $2\leqslant\ell\leqslant300$ for the former, while  $30\leqslant\ell\leqslant300$ for the latter. When we estimate the uncertainties on $r$ and $n_t$ simultaneously, we only consider the sky coverage $f_{\textrm{sky}}=72\%$.

\subsection{Uncertainty on $r$}
\begin{figure*}
\begin{minipage}[t]{\columnwidth}
\centering
\includegraphics[width=\columnwidth]{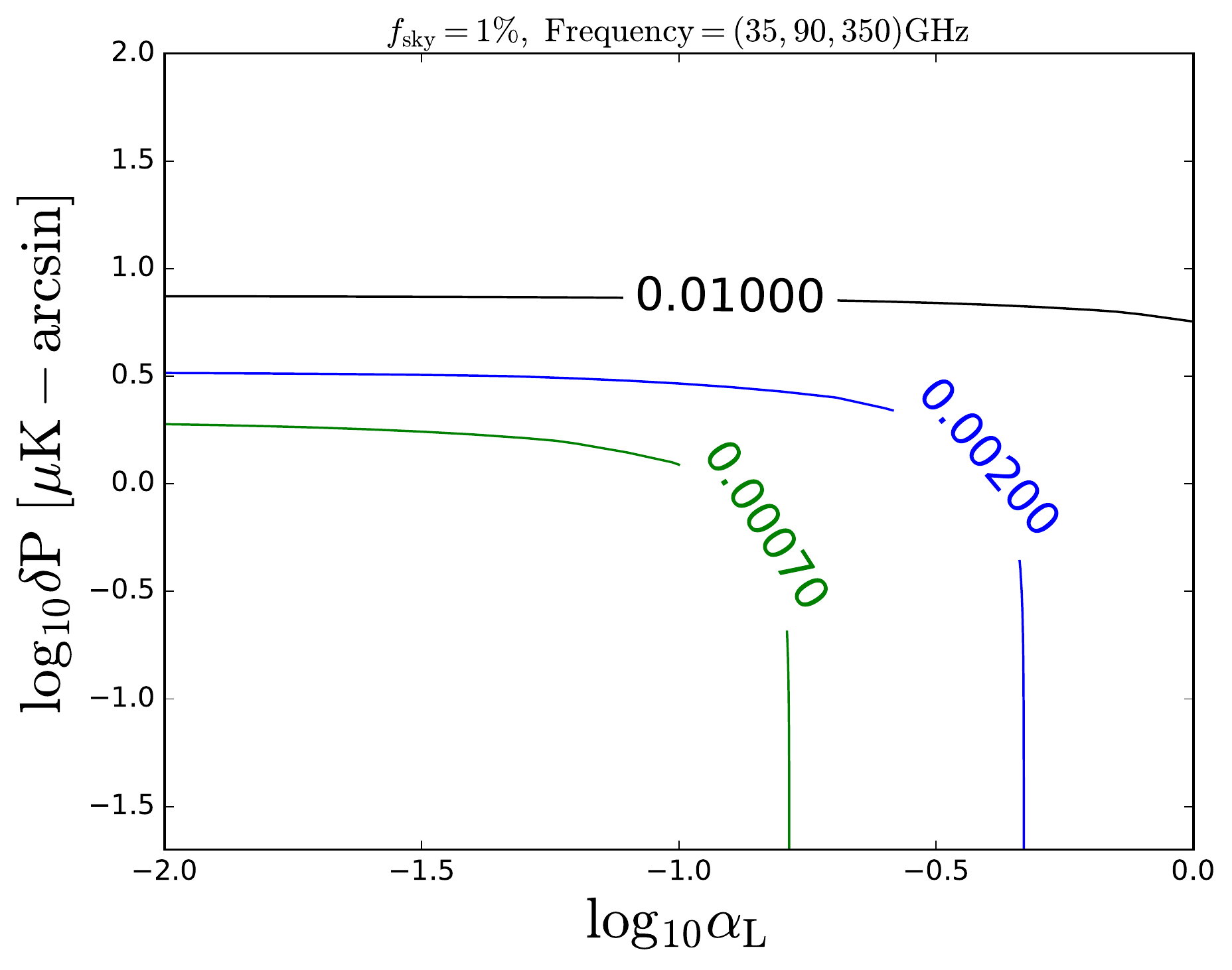}
\vspace{0.1cm}
\end{minipage}\hspace*{0.1\columnwidth}%
\begin{minipage}[t]{\columnwidth}
\centering
\includegraphics[width=\columnwidth]{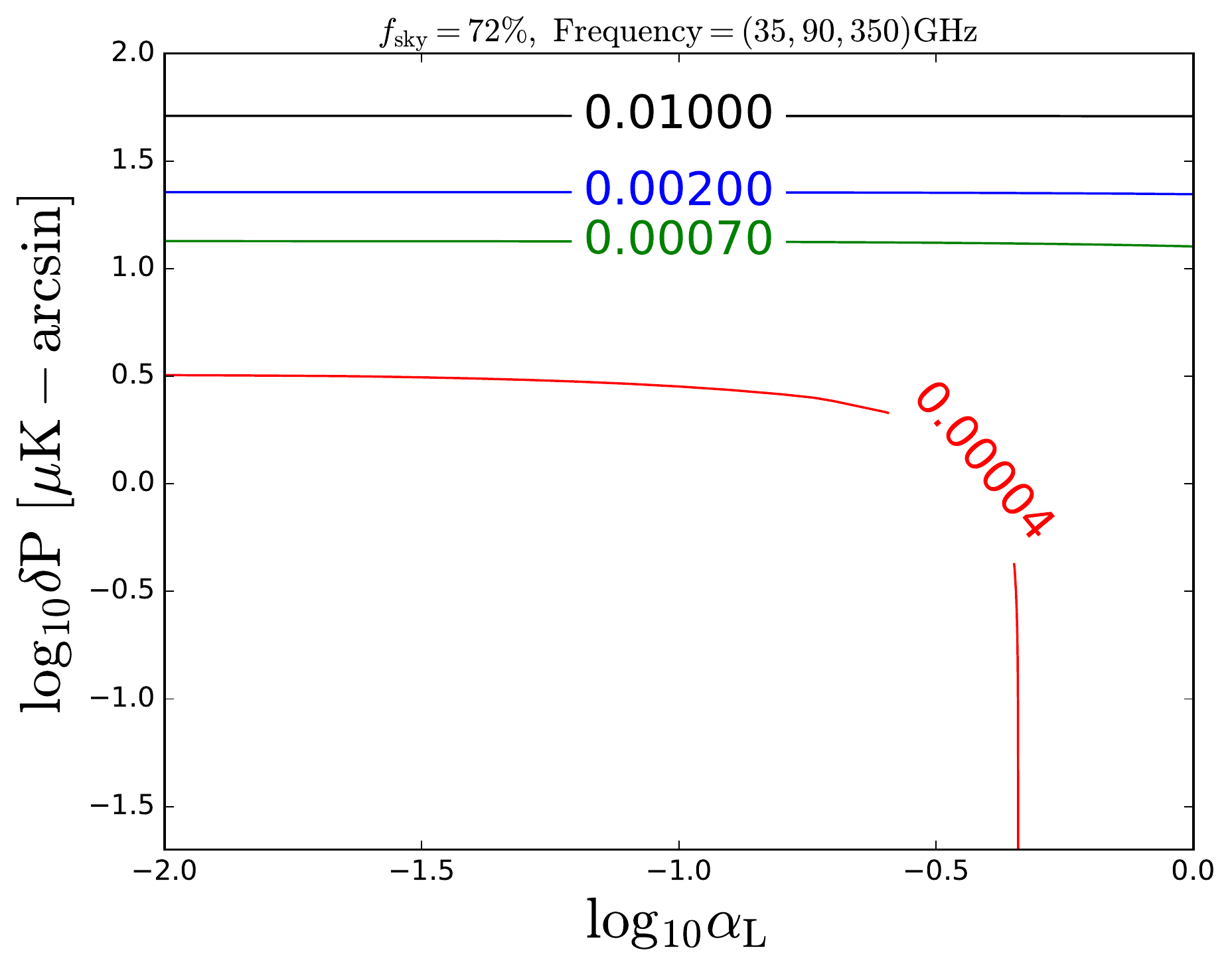}
\vspace{0.1cm}
\end{minipage}
\begin{minipage}[t]{\columnwidth}
\centering
\includegraphics[width=\columnwidth]{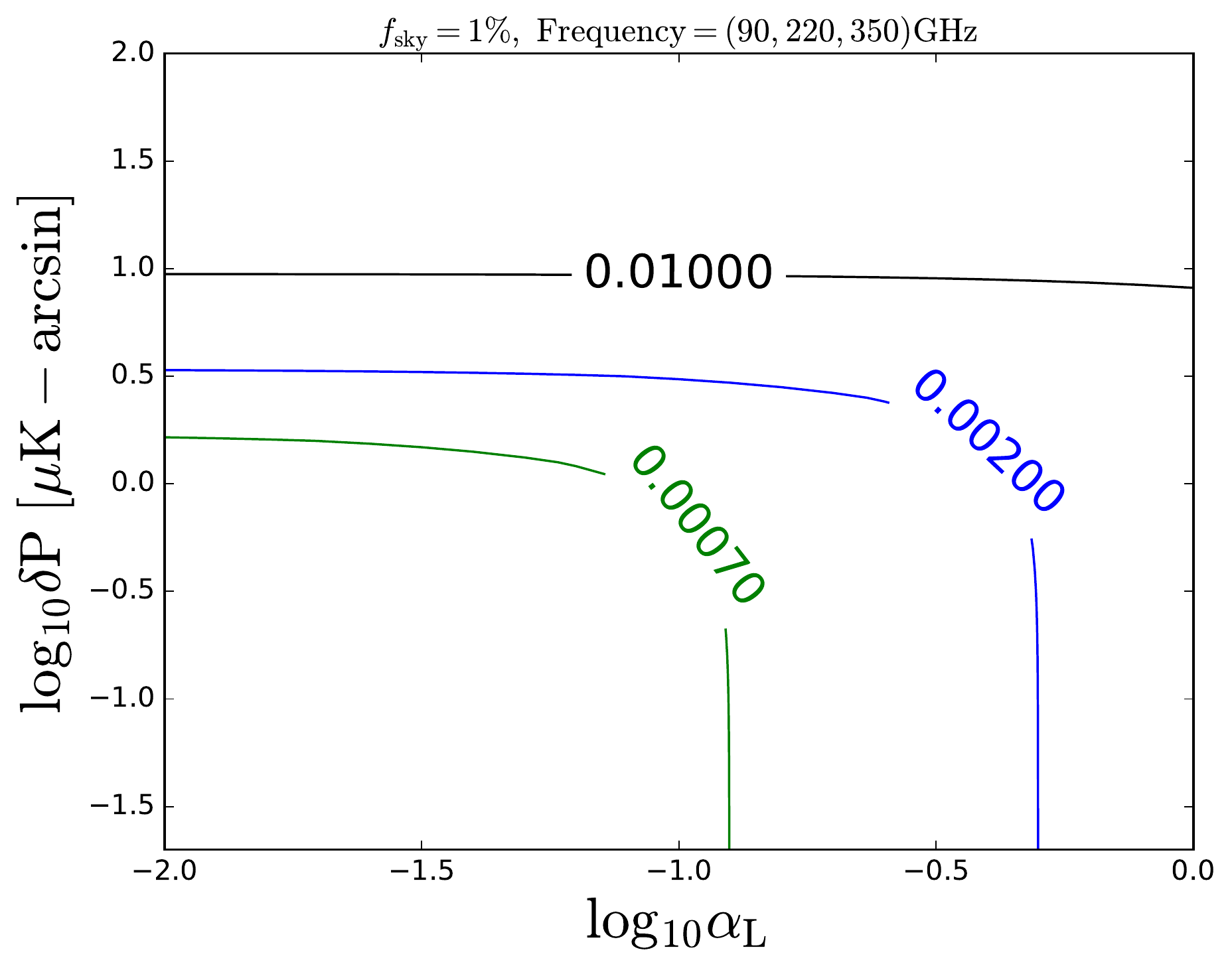}
\vspace{0.1cm}
\end{minipage}\hspace*{0.1\columnwidth}%
\begin{minipage}[t]{\columnwidth}
\centering
\includegraphics[width=\columnwidth]{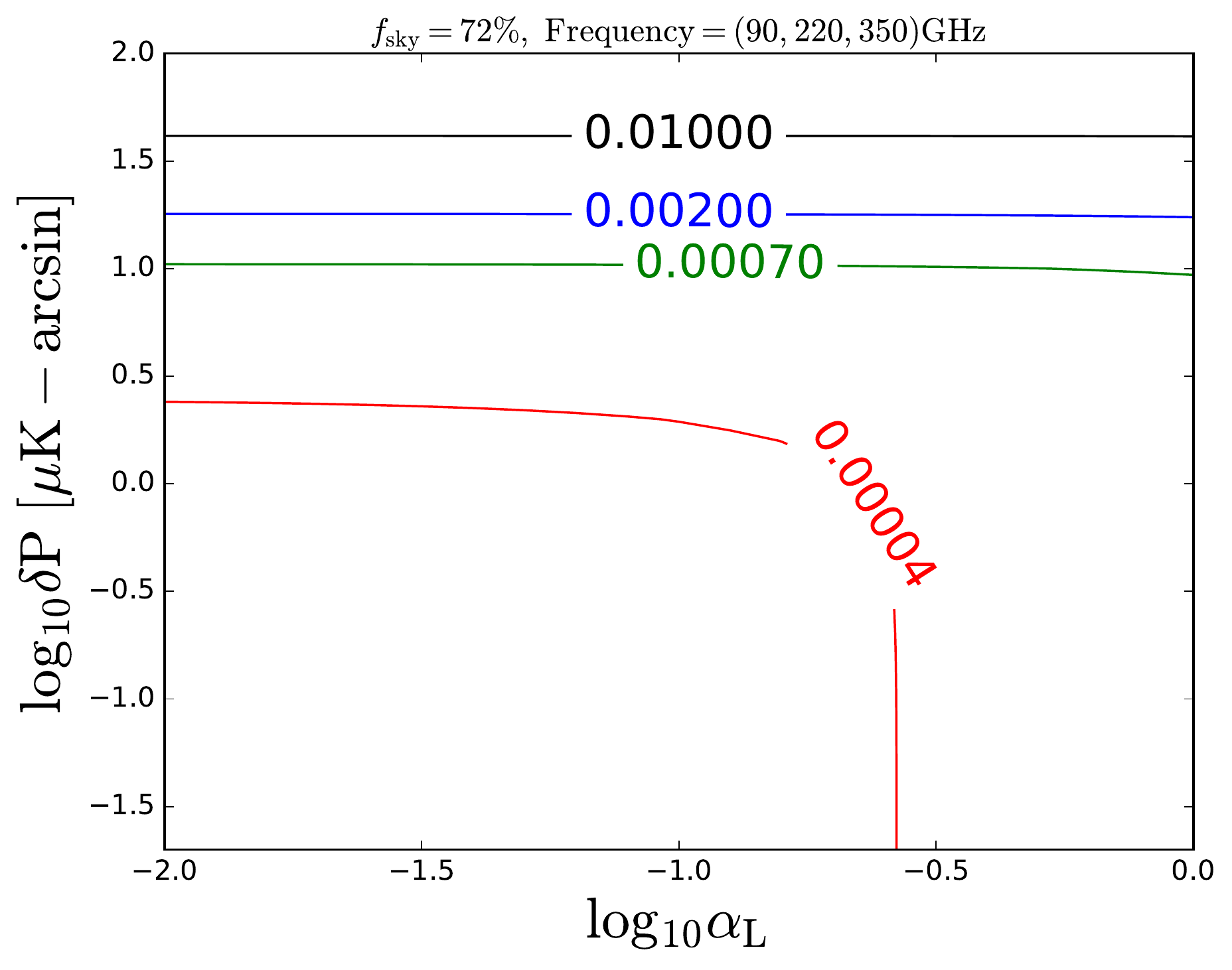}
\vspace{0.1cm}
\end{minipage}
\begin{minipage}[t]{\columnwidth}
\centering
\includegraphics[width=\columnwidth]{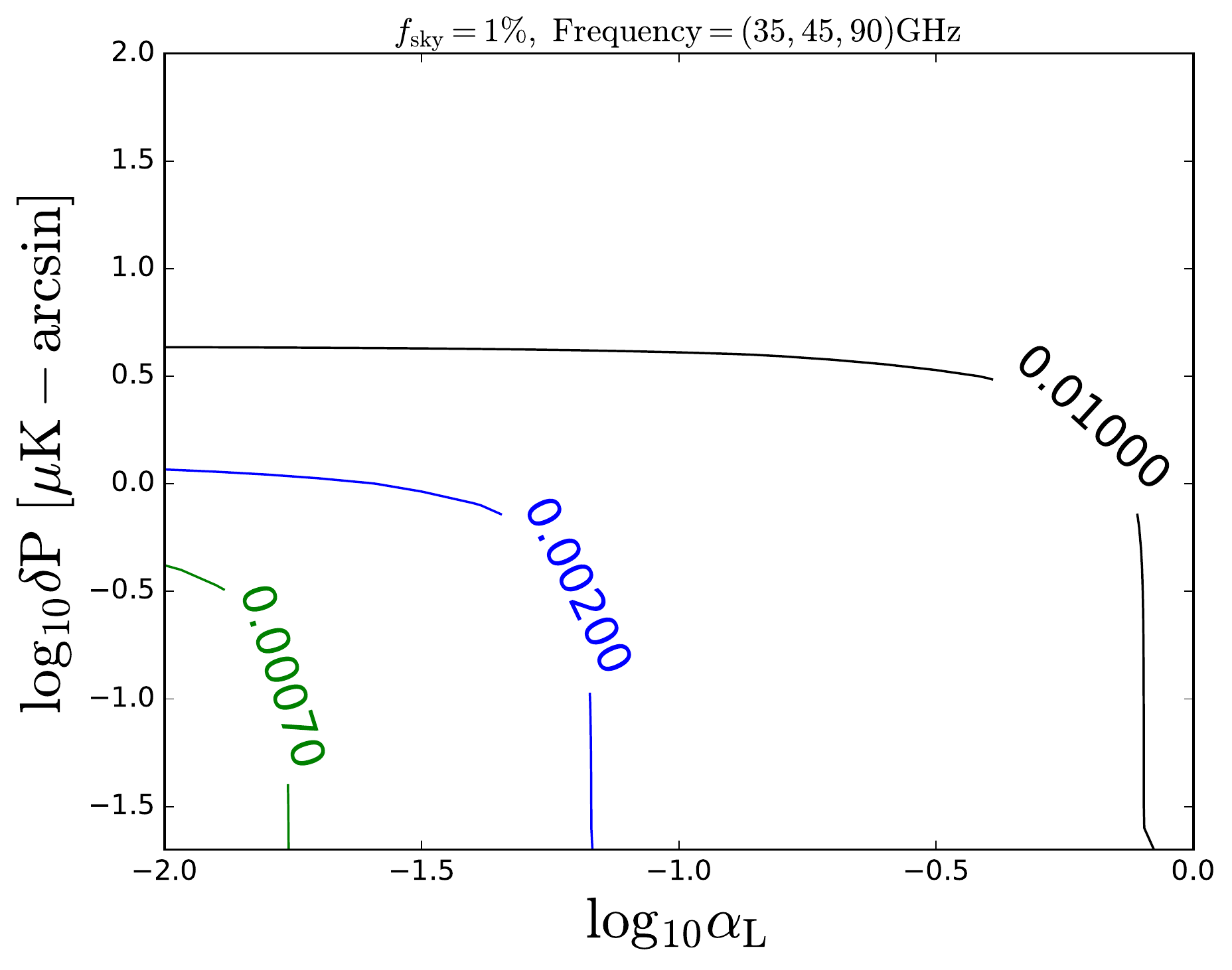}
\end{minipage}\hspace*{0.1\columnwidth}%
\begin{minipage}[t]{\columnwidth}
\centering
\includegraphics[width=\columnwidth]{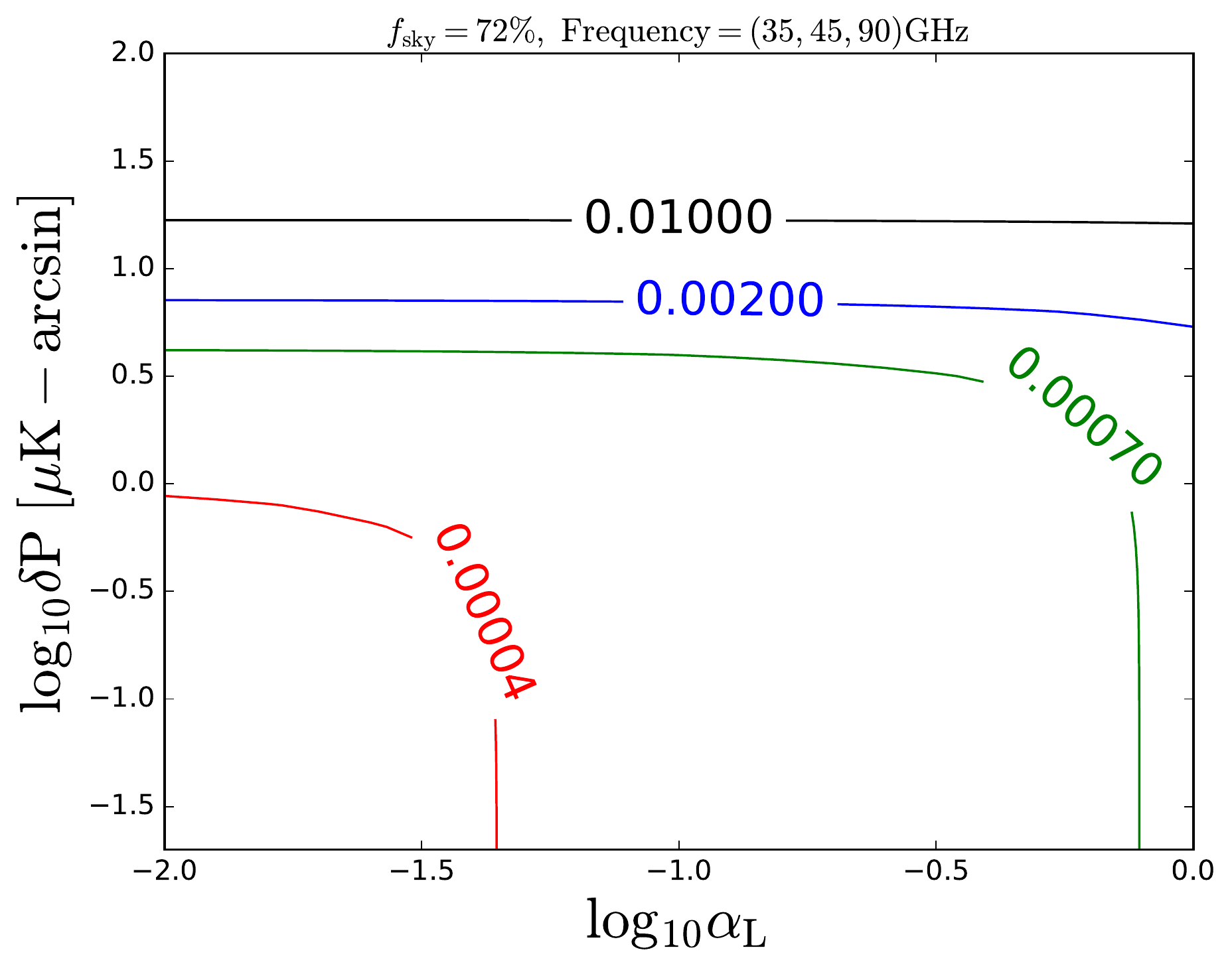}
\end{minipage}
\caption{\small The $1\sigma$ uncertainty $\sigma_r$ in the ($\alpha_L$, $\delta P$) plane for ($35$, $90$, $350$)~GHz (upper), ($90$, $220$, $350$)~GHz (middle) and ($35$, $45$, $90$)~GHz (lower) when $f_{\textrm{sky}}=1\%$ (left) and $72\%$ (right). We set $\bar{r}=0$ as the fiducial model. The red curve denotes a contour of $\sigma_r=4\times10^{-5}$, the green one $\sigma_r=7\times10^{-4}$, the blue one $\sigma_r=2\times10^{-3}$ and the black one $\sigma_r=10^{-2}$.}
\label{fig:sigmar}
\end{figure*}

\noindent
The polarized dust foreground has crucial influence on the $r$ constraints \citep{Creminelli:2015oda}. Taking the polarized dust into account, we wonder whether an improvement of $\delta P$ and $\alpha_L$ can have potential impacts on constraining $r$. 
We denote the $1\sigma$ uncertainty on $r$ with $\sigma_{r}$.
Fig.~\ref{fig:sigmar} shows the variation of $\sigma_{r}$ as the noise and delensing parameters are varied.
The left subfigures are referred to $f_{\textrm{sky}}=1\%$, while the right ones $f_{\textrm{sky}}=72\%$. The top subfigures are showed for the frequency bands (35, 90, 350) GHz, the middle ones for (90, 220, 350) GHz, and the bottom ones for (35, 45, 90) GHz. In each subfigure, we depict four typical contours of $\sigma_r$, namely, the red curve curve for $\sigma_r=4\times10^{-5}$ (no display for $f_{\rm{sky}}=1\%$), the green one for $\sigma_r=7\times10^{-4}$, the blue one for $\sigma_r=2\times10^{-3}$, and the black one for $\sigma_r=10^{-2}$.

From Fig.~\ref{fig:sigmar}, we find that $\delta P$ and $\alpha_L$ can significantly impact the uncertainty on $r$ for a vanishing fiducial $r$. 
Decreasing the instrumental noise always leads to better constraints on $r$.
For $f_{\textrm{sky}}=72\%$, the low-$\ell$ multipoles of the CMB B-modes are considered, which covers the reionization peak. Thus the delensing becomes essential when $r$ is lower than the order of $10^{-3}$. By contrast, the low-$\ell$ multipoles of the CMB B-modes are not taken into account for $f_{\textrm{sky}}=1\%$. Thus the delensing becomes important only when $r$ is lower than the order $10^{-2}$. In fact, it is marginally important to take delensing when $r\gtrsim \rm{few}\times10^{-2}$.

For $f_{\textrm{sky}}=72\%$, all three experimental setups can reach the sensitivity $\sigma_r=4\times10^{-5}$. In particular, the frequency combination (35, 90, 350) GHz can even reach $\sigma_r\sim \mathcal{O}(10^{-6})$. This setup covers the synchrotron band, the CMB band and the polarized dust band simultaneously. If the synchrotron band is discarded, the sensitivity becomes lower by around two times. If the polarized dust band is discarded, the sensitivity becomes lower by about one order. Our results reveal that the polarized dust foreground has more significant impact on a probe of the primordial B-modes than the synchrotron foreground does. For $f_{\textrm{sky}}=1\%$, by contrast, the sensitivity on probing $r$ becomes much lower for all three experimental setups, due to the large uncertainty on the index $\beta_D$ of the polarized dust foreground and the absence of the low-$\ell$ multipoles of the CMB B-modes. However, the sensitivity $\sigma_r=7\times10^{-4}$ may be reachable. In this case, the most sensitive setup is still given by (35, 90, 350) GHz.

\subsection{Uncertainty on $n_t$}
\begin{figure}
\centering
\begin{minipage}[t]{\columnwidth}
\centering
\includegraphics[width=\columnwidth]{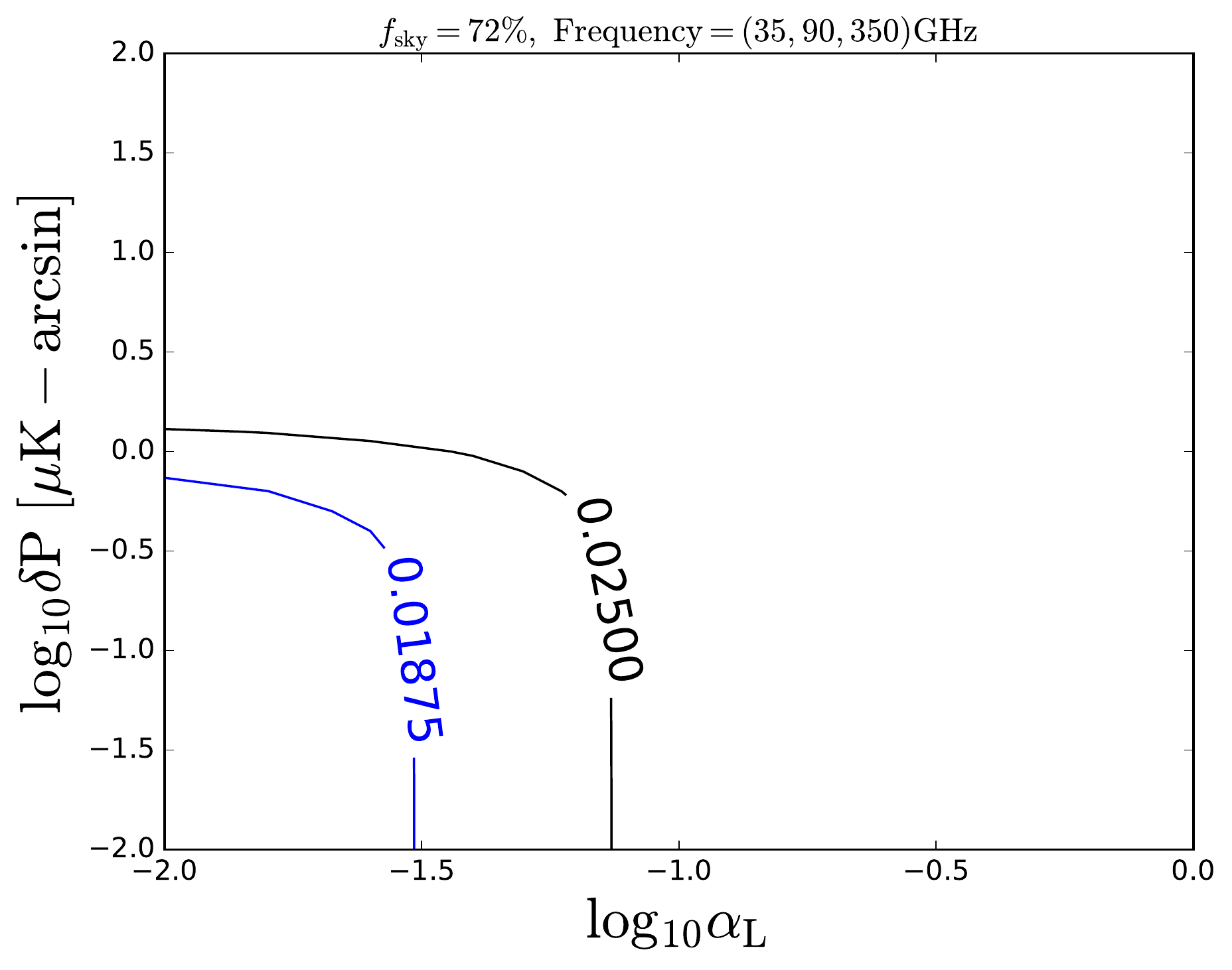}
\vspace{0.1cm}
\end{minipage}
\begin{minipage}[t]{\columnwidth}
\centering
\includegraphics[width=\columnwidth]{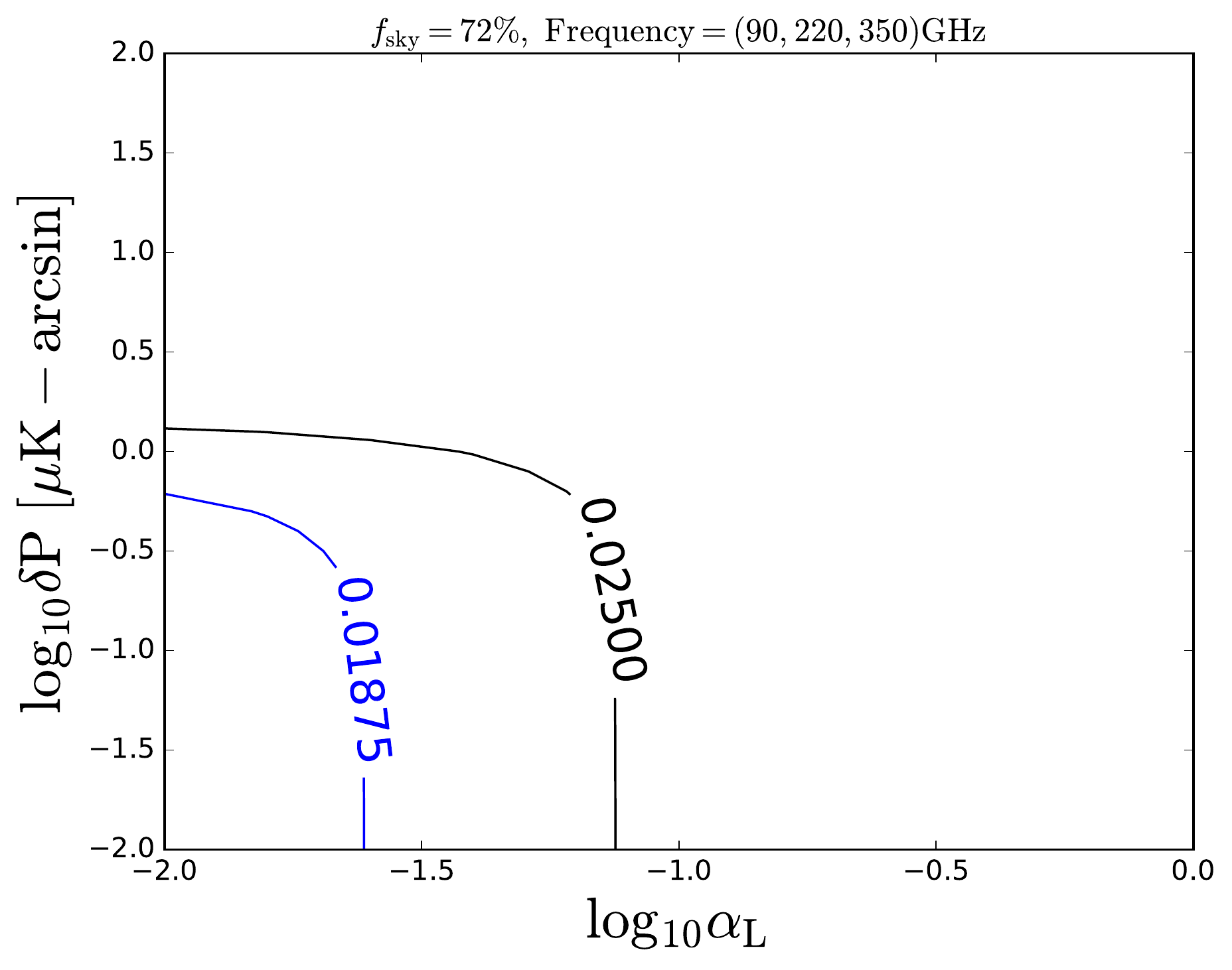}
\vspace{0.1cm}
\end{minipage}
\begin{minipage}[t]{\columnwidth}
\centering
\includegraphics[width=\columnwidth]{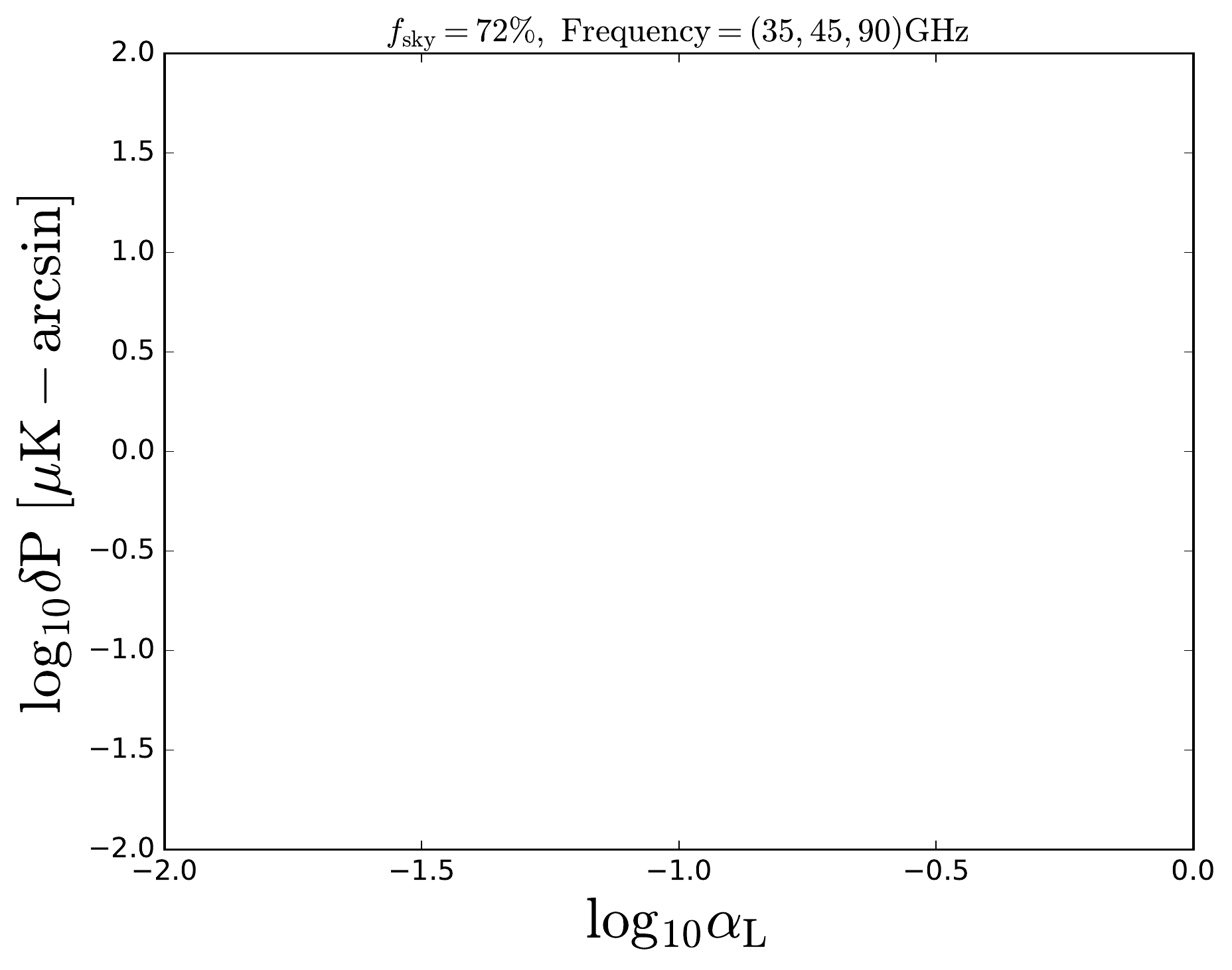}
\end{minipage}
\caption{\small The $1\sigma$ uncertainty $\sigma_{n_t}$ in the ($\alpha_L$, $\delta P$) plane for ($35$, $90$, $350$)~GHz (upper), ($90$, $220$, $350$)~GHz (middle) and ($35$, $45$, $90$)~GHz (lower) when $f_{\textrm{sky}}=72\%$. We set $\bar{r}=0.05$ and $\bar{n_t}=-\bar{r}/8$ as the fiducial model. The red curve denotes a contour of $\sigma_{n_t}=\bar{r}/8$, the green one $\sigma_{n_t}=2\bar{r}/8$, the blue one $\sigma_{n_t}=3\bar{r}/8$ and the black one $\sigma_{n_t}=4\bar{r}/8$. Please note that the red and green curves are outside of the plot (in the lower left corner). For the lower panel, all the curves are outside of the plot (in the lower left corner).}
\label{fig:sigmarnt}
\end{figure}

\noindent
The polarized dust radiations have significant impacts on the $n_t$ constraints \citep{Huang:2015gca}. For the optimistic consideration, we wonder how precisely one can measure the power spectrum of primordial B-modes and if one can distinguish it from the exactly scale-invariant spectrum. In the ($\alpha_L$, $\delta P$) plane, the two-dimensional contours for the $1\sigma$ uncertainty on $n_t$ are depicted in Fig.~\ref{fig:sigmarnt}. 
The top subfigure is showed for (35, 90, 350) GHz, the middle one for (90, 220, 350) GHz, and the bottom one for (35, 45, 90) GHz. As mentioned in Subsec.~\ref{subsec:fisher}, we have fixed $\bar{r}=0.05$ here. In each subfigure, we depict four typical contours of $\sigma_{n_t}$, namely, the red curve for $\sigma_{n_t}=\bar{r}/8$, the green one for $\sigma_{n_t}=\bar{r}/4$, the blue one for $\sigma_{n_t}=3\bar{r}/8$, and the black one for $\sigma_{n_t}=\bar{r}/2$. 
However, the red and green curves do not be displayed here (in the lower left corner of panels) since the required sensitivities have not been reached. In the lower panel, all the curves are outside of the plot (in the lower left corner).

To be optimistic, we reach the best sensitivity $\sigma_{n_t}\simeq0.015$. This is still larger than the cosmic-variance limit showed in the next subsection. Given the current constraint $r<0.07$ at $95\%$ CL \citep{Array:2015xqh,Huang:2015cke}, therefore, one can not discriminate the consistency relation $n_t=-r/8$ from the scale-invariant spectrum, even for the most optimistic scenario in this paper.
However, the observations of CMB B-mode polarization can still place stringent constraints on $n_t$. 
Similar to $r$ in Fig.~\ref{fig:sigmar}, the sensitivity to $n_t$ can be impacted significantly by both $\delta P$ and $\alpha_L$ according to Fig.~\ref{fig:sigmarnt}. The setup  (35, 90, 350) GHz and (90, 220, 350) GHz provide the best sensitivity to $n_t$. The reason is that these setup can determine the parameters $A_D$ and $\beta_D$ of the polarized dust foreground with higher precision. Therefore, a multi-band observation of the CMB B-mode polarization, determining the polarized dust emission better, may be helpful to learn more about $n_t$.



\subsection{Cosmic-Variance Limit on $n_t$}
\noindent
The cosmic variance (CV), i.e. $\Delta C_{\ell}/C_{\ell}=\sqrt{2/[(2\ell+1)f_{\rm{sky}}]}$ for the $\ell$-th multipole, is inevitable in the CMB power spectra. It limits the precision in determining cosmological parameters. Due to the CV limit, for example, one cannot constrain $r$ better than $0.05$ with the CMB temperature anisotropies only (this amplitude will be improved by four times with the E-mode polarization only) \footnote{http://cosmologist.info/notes/tensors.ps}. Given the present stringent constraints on $r$, future improvements will mainly come from the non-CV-limited B-mode polarization of the CMB. However, the CV of the B-mode power spectrum still limits the precision in measuring $r$ with the CMB B-mode polarization only. It can thus give rise to the CV limit on the determination of $n_t$. We will study this CV limit on $n_{t}$ in the following.

\begin{figure}
\centering
\includegraphics[width=\columnwidth]{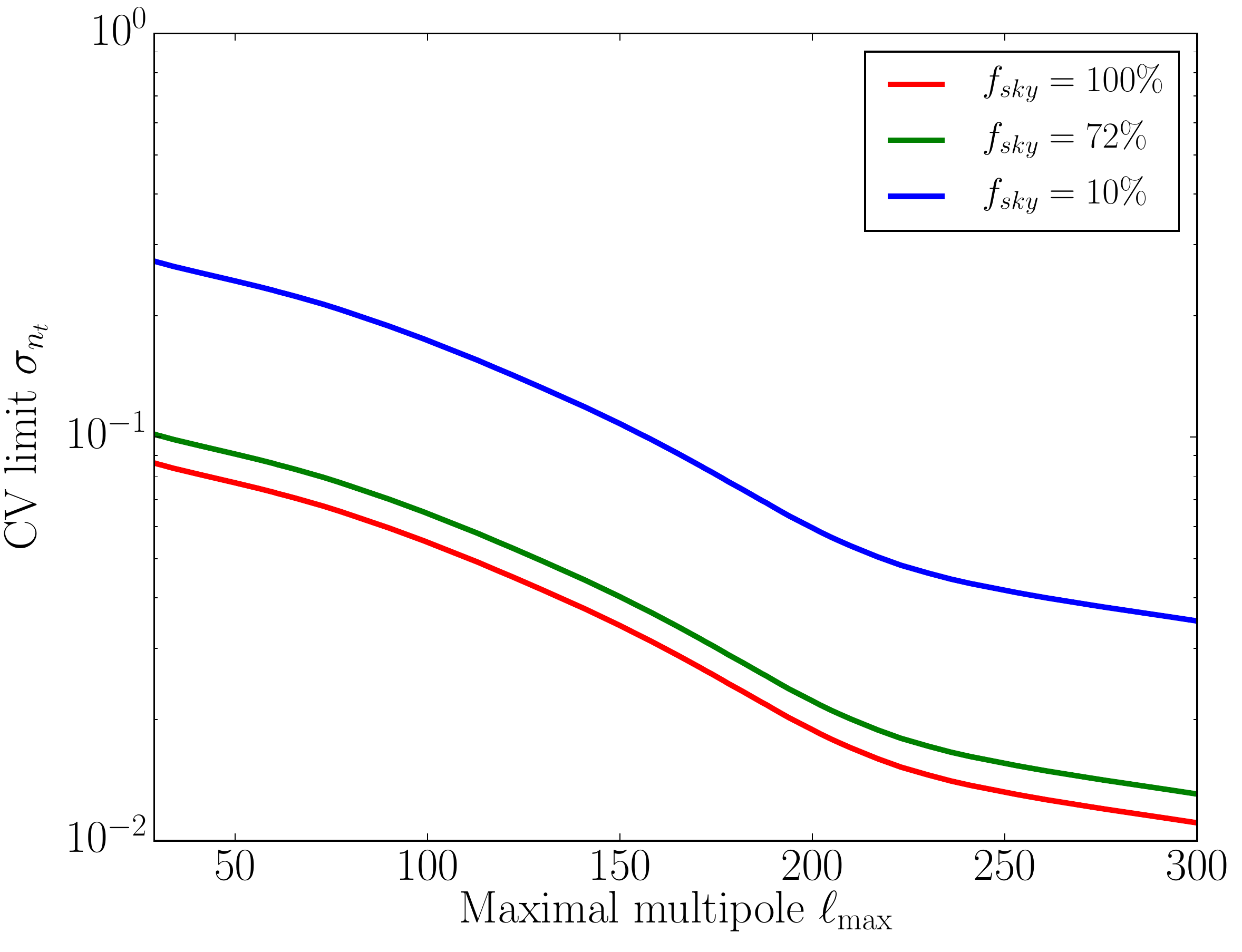}
\caption{\small The CV limit on $n_t$ versus the maximal multipole $\ell_{\rm{max}}$. For a given value of $\ell_{\rm{max}}$, we compute the $1\sigma$ uncertainty $\sigma_{n_t}$ with the multipoles of CMB B-modes from $2$ to $\ell_{\rm{max}}$.}
\label{fig:cvlimitnt}
\end{figure}

To find the CV limit on $n_t$, we consider only the primordial B-mode polarization of the CMB, regardless of the contamination from foregrounds, noise and lensing. Therefore, the average log-likelihood in Eq.~(\ref{likelihood}) becomes
\begin{equation}
\label{cvlikelihood}
\langle\ln \mathcal{L}_{BB}\rangle=-\frac{1}{2}\sum_{\ell=2}^{\ell_{\rm{max}}}f_{\rm{sky}}(2\ell+1)\left(\ln \tilde{C}_{\ell}+\frac{\bar{\tilde{C}}_{\ell}}{\tilde{C}_{\ell}}\right)\ ,
\end{equation}
where only one frequency band is considered. Calculating the Fisher information matrix, we show the CV limit on $n_t$ as a function of the maximal multipole $\ell_{\rm{max}}$ in Fig.~\ref{fig:cvlimitnt}. 
For each given value of $\ell_{\rm{max}}$, we compute the $1\sigma$ uncertainty $\sigma_{n_t}$ using the CMB multipoles of B-mode polarization from $2$ to $\ell_{\rm{max}}$. We choose $29\leqslant\ell_{\rm{max}}\leqslant300$ in this study. In principle, the CV limits for higher $\ell_{\rm{max}}$ can be obtained in the same way. Obviously, an extension to higher $\ell_{\rm{max}}$ can improve the CV limit on $n_t$. However, this is not necessary in the near future, since the contamination from the residual power of the lensing B-mode polarization is more dominant for higher multipoles (please refer to Fig.~\ref{fig:lensr}).

The CV limit on $n_t$ only depends on the CV of the CMB B-mode power spectrum, i.e., $\Delta C_{\ell}^{\rm{BB}}/C_{\ell}^{\rm{BB}}=\sqrt{2/[(2\ell+1)f_{\rm{sky}}]}$. The B-mode power thus preserves a smaller uncertainty for a higher $\ell$. Therefore, $\sigma_{n_t}$ decreases monotonically with $\ell_{\rm{max}}$ in Fig.~\ref{fig:cvlimitnt}. For example, in the case of full sky, our results show $\sigma_{n_t}\simeq0.09$ for $2\leqslant\ell\leqslant29$ (low-$\ell$ multipoles) while $\sigma_{n_t}\simeq0.01$ for $2\leqslant\ell\leqslant300$. In addition, $\sigma_{n_t}$ becomes smaller for a larger sky coverage which covers many more multipoles of the CMB B-mode polarization. According to Fig.~\ref{fig:cvlimitnt}, for example, $\sigma_{n_t}$ for the full sky is smaller by around three times than that for the $10\%$ sky. In addition, we study the fiducial model with $\bar{n}_{t}=0$, for which the CV limit on $n_t$ does not depend on the fiducial value of $r$. We find the same CV-limit on $n_t$.

Given the current constraint $r<0.07$ at $95\%$ CL \citep{Array:2015xqh,Huang:2015cke}, it is unlikely to distinguish the consistency relation $n_t=-r/8$ from the scale invariance, i.e. $n_t=0$, by using the CMB B-mode only. The full-sky CV limit on $n_t$, denoted by the red curve in Fig.~\ref{fig:cvlimitnt}, cannot be crossed over with the CMB B-mode observations only. Therefore, $n_t=-r/8$ would be probably consistent with the scale invariance within $1\sigma$ CL. To further reduce the uncertainty on $n_t$, one needs to supplement the CMB data with other external datasets \footnote{http://vega.ess.sci.osaka-u.ac.jp/seminar/semiold/files/\\20130902Hazumi.pdf}. For example, one may test the consistency relation at $2\sigma$ CL using the 21cm line \citep{Masui:2010cz}. However, future observations on the CMB polarizations can still make relatively stringent constraints on $n_t$ \citep{Huang:2015gca}, and have potentials to rule out some models of the very early Universe, for example, the ekpyrotic universe model which predicts $n_t=2$ \citep{Khoury:2001wf}.

\section{Conclusion and Discussion}
\label{sec:conclusion}
\noindent
In this paper, we provided optimistic estimations on the sensitivity to the detection of the primordial gravitational waves using the CMB B-mode polarization only. In the measurements of $r$ with three frequency bands, we found the optimistic setup to be (35, 90, 350) GHz, which covers the synchrotron, CMB, and polarized dust bands simultaneously. In the most optimistic scenario, the sensitivity $\sigma_r\simeq4\times10^{-5}$ may be reachable for the $72\%$ sky coverage. This sensitivity is decreased by about one order for the $1\%$ sky coverage. In the measurements of $n_t$, we found that the best sensitivity is $\sigma_{n_t}\simeq0.015$ for the same optimistic setup. Our results confirm the significant impact of the polarized dust on the observations of the primordial B-modes. The conclusions in this work are consistent with our previous estimation in Reference \citep{Huang:2015gca}.

The sensitivity to $n_t$ is inevitably limited by the cosmic variance of the power spectrum of CMB B-mode polarization. In this paper, we have estimated this CV limit on the measurements of $n_t$. Given the fiducial model $\bar{r}=0.05$ and $\bar{n}_{t}=-\bar{r}/8$, the CV limit on $n_{t}$ is $\sigma_{n_t}\simeq0.01$ for $2\leqslant\ell\leqslant300$. This is expected to be improved by including higher-order multipoles (i.e. $\ell>300$) of the CMB B-mode polarization. If we use the fiducial model with $\bar{n}_{t}=0$, the CV limit on $n_{t}$ is independent of $\bar{r}$. We found that it is $\sigma_{n_t}\simeq0.01$ for $2\leqslant\ell\leqslant300$. Given the up-to-date constraint $r_{0.05}<0.07$ at 95\% CL, the consistency relation $n_t=-r/8$ cannot be distinguished from the scale-invariant spectrum using the CMB B-mode polarization only. Even though challenging, however, it is still possible to discriminate some models of the very early Universe, e.g. the ekpyrotic universe model, by only observing the CMB B-mode polarization in the future.


\section*{Acknowledgements}
\noindent We acknowledge the use of HPC Cluster of SKLTP/ITP-CAS.
Q.-G.H. is supported by grants from NSFC 
(grant No. 11335012, 11575271, 11690021, 11747601), 
the Strategic Priority Research Program of Chinese Academy of Sciences 
(Grant No. XDB23000000), Top-Notch Young Talents Program of China, 
and Key Research Program of Frontier Sciences of CAS. 
S.W. is partially supported by funding from the Research Grants Council of the Hong Kong Special Administrative Region, China (Project No. 14301214). 



\bibliographystyle{mnras}
\bibliography{myreference} 

\begin{thebibliography}{}
\makeatletter
\relax
\def\mn@urlcharsother{\let\do\@makeother \do\$\do\&\do\#\do\^\do\_\do\%\do\~}
\def\mn@doi{\begingroup\mn@urlcharsother \@ifnextchar [ {\mn@doi@}
  {\mn@doi@[]}}
\def\mn@doi@[#1]#2{\def\@tempa{#1}\ifx\@tempa\@empty \href
  {http://dx.doi.org/#2} {doi:#2}\else \href {http://dx.doi.org/#2} {#1}\fi
  \endgroup}
\def\mn@eprint#1#2{\mn@eprint@#1:#2::\@nil}
\def\mn@eprint@arXiv#1{\href {http://arxiv.org/abs/#1} {{\tt arXiv:#1}}}
\def\mn@eprint@dblp#1{\href {http://dblp.uni-trier.de/rec/bibtex/#1.xml}
  {dblp:#1}}
\def\mn@eprint@#1:#2:#3:#4\@nil{\def\@tempa {#1}\def\@tempb {#2}\def\@tempc
  {#3}\ifx \@tempc \@empty \let \@tempc \@tempb \let \@tempb \@tempa \fi \ifx
  \@tempb \@empty \def\@tempb {arXiv}\fi \@ifundefined
  {mn@eprint@\@tempb}{\@tempb:\@tempc}{\expandafter \expandafter \csname
  mn@eprint@\@tempb\endcsname \expandafter{\@tempc}}}

\bibitem[\protect\citeauthoryear{Adam et~al.}{Adam
  et~al.}{2016a}]{Adam:2014bub}
Adam R.,  et~al., 2016a, \mn@doi [Astron. Astrophys.]
  {10.1051/0004-6361/201425034}, 586, A133

\bibitem[\protect\citeauthoryear{Adam et~al.}{Adam
  et~al.}{2016b}]{Adam:2016hgk}
Adam R.,  et~al., 2016b, \mn@doi [Astron. Astrophys.]
  {10.1051/0004-6361/201628897}, 596, A108

\bibitem[\protect\citeauthoryear{Ade et~al.}{Ade et~al.}{2015}]{Ade:2015tva}
Ade P. A.~R.,  et~al., 2015, \mn@doi [Phys. Rev. Lett.]
  {10.1103/PhysRevLett.114.101301}, 114, 101301

\bibitem[\protect\citeauthoryear{Ade et~al.}{Ade et~al.}{2016a}]{Array:2015xqh}
Ade P. A.~R.,  et~al., 2016a, \mn@doi [Phys. Rev. Lett.]
  {10.1103/PhysRevLett.116.031302}, 116, 031302

\bibitem[\protect\citeauthoryear{Ade et~al.}{Ade et~al.}{2016b}]{Ade:2015xua}
Ade P. A.~R.,  et~al., 2016b, \mn@doi [Astron. Astrophys.]
  {10.1051/0004-6361/201525830}, 594, A13

\bibitem[\protect\citeauthoryear{Ade et~al.}{Ade et~al.}{2016c}]{Ade:2015lrj}
Ade P. A.~R.,  et~al., 2016c, \mn@doi [Astron. Astrophys.]
  {10.1051/0004-6361/201525898}, 594, A20

\bibitem[\protect\citeauthoryear{Ade et~al.}{Ade et~al.}{2016d}]{Ade:2015nch}
Ade P. A.~R.,  et~al., 2016d, \mn@doi [Astron. Astrophys.]
  {10.1051/0004-6361/201527932}, 596, A102

\bibitem[\protect\citeauthoryear{Ade et~al.}{Ade et~al.}{2018}]{Ade:2018gkx}
Ade P. A.~R.,  et~al., 2018, Submitted to: Phys. Rev. Lett.

\bibitem[\protect\citeauthoryear{Aghanim et~al.}{Aghanim
  et~al.}{2017}]{Aghanim:2016cps}
Aghanim N.,  et~al., 2017, \mn@doi [Astron. Astrophys.]
  {10.1051/0004-6361/201629164}, 599, A51

\bibitem[\protect\citeauthoryear{Albrecht \& Steinhardt}{Albrecht \&
  Steinhardt}{1982}]{Albrecht:1982wi}
Albrecht A.,  Steinhardt P.~J.,  1982, \mn@doi [Phys. Rev. Lett.]
  {10.1103/PhysRevLett.48.1220}, 48, 1220

\bibitem[\protect\citeauthoryear{Baumann, Lee  \& Pimentel}{Baumann
  et~al.}{2016}]{Baumann:2015xxa}
Baumann D.,  Lee H.,   Pimentel G.~L.,  2016, \mn@doi [JHEP]
  {10.1007/JHEP01(2016)101}, 01, 101

\bibitem[\protect\citeauthoryear{Cabass, Pagano, Salvati, Gerbino, Giusarma  \&
  Melchiorri}{Cabass et~al.}{2016}]{Cabass:2015jwe}
Cabass G.,  Pagano L.,  Salvati L.,  Gerbino M.,  Giusarma E.,   Melchiorri A.,
   2016, \mn@doi [Phys. Rev.] {10.1103/PhysRevD.93.063508}, D93, 063508

\bibitem[\protect\citeauthoryear{Cheng, Huang  \& Wang}{Cheng
  et~al.}{2014}]{Cheng:2014pxa}
Cheng C.,  Huang Q.-G.,   Wang S.,  2014, \mn@doi [JCAP]
  {10.1088/1475-7516/2014/12/044}, 1412, 044

\bibitem[\protect\citeauthoryear{Creminelli, Gleyzes, Nore?a  \&
  Vernizzi}{Creminelli et~al.}{2014}]{Creminelli:2014wna}
Creminelli P.,  Gleyzes J.,  Nore?a J.,   Vernizzi F.,  2014, \mn@doi [Phys.
  Rev. Lett.] {10.1103/PhysRevLett.113.231301}, 113, 231301

\bibitem[\protect\citeauthoryear{Creminelli, Lopez~Nacir, Simonovic, Trevisan
  \& Zaldarriaga}{Creminelli et~al.}{2015}]{Creminelli:2015oda}
Creminelli P.,  Lopez~Nacir D.~L.,  Simonovic M.,  Trevisan G.,   Zaldarriaga
  M.,  2015, \mn@doi [JCAP] {10.1088/1475-7516/2015/11/031}, 1511, 031

\bibitem[\protect\citeauthoryear{Crittenden, Bond, Davis, Efstathiou  \&
  Steinhardt}{Crittenden et~al.}{1993}]{Crittenden:1993ni}
Crittenden R.,  Bond J.~R.,  Davis R.~L.,  Efstathiou G.,   Steinhardt P.~J.,
  1993, \mn@doi [Phys. Rev. Lett.] {10.1103/PhysRevLett.71.324}, 71, 324

\bibitem[\protect\citeauthoryear{Errard, Feeney, Peiris  \& Jaffe}{Errard
  et~al.}{2016}]{Errard:2015cxa}
Errard J.,  Feeney S.~M.,  Peiris H.~V.,   Jaffe A.~H.,  2016, \mn@doi [JCAP]
  {10.1088/1475-7516/2016/03/052}, 1603, 052

\bibitem[\protect\citeauthoryear{Escudero, Ramirez, Boubekeur, Giusarma  \&
  Mena}{Escudero et~al.}{2016}]{Escudero:2015wba}
Escudero M.,  Ramirez H.,  Boubekeur L.,  Giusarma E.,   Mena O.,  2016,
  \mn@doi [JCAP] {10.1088/1475-7516/2016/02/020}, 1602, 020

\bibitem[\protect\citeauthoryear{Flauger, Hill  \& Spergel}{Flauger
  et~al.}{2014}]{Flauger:2014qra}
Flauger R.,  Hill J.~C.,   Spergel D.~N.,  2014, \mn@doi [JCAP]
  {10.1088/1475-7516/2014/08/039}, 1408, 039

\bibitem[\protect\citeauthoryear{Garcia-Bellido, Roest, Scalisi  \&
  Zavala}{Garcia-Bellido et~al.}{2014}]{Garcia-Bellido:2014wfa}
Garcia-Bellido J.,  Roest D.,  Scalisi M.,   Zavala I.,  2014, \mn@doi [Phys.
  Rev.] {10.1103/PhysRevD.90.123539}, D90, 123539

\bibitem[\protect\citeauthoryear{Grishchuk}{Grishchuk}{1975}]{Grishchuk:1974ny}
Grishchuk L.,  1975, Soviet Journal of Experimental and Theoretical Physics,
  40, 409

\bibitem[\protect\citeauthoryear{Guth}{Guth}{1981}]{Guth:1980zm}
Guth A.~H.,  1981, \mn@doi [Phys. Rev.] {10.1103/PhysRevD.23.347}, D23, 347

\bibitem[\protect\citeauthoryear{Guzzetti, Bartolo, Liguori  \&
  Matarrese}{Guzzetti et~al.}{2016}]{Guzzetti:2016mkm}
Guzzetti M.~C.,  Bartolo N.,  Liguori M.,   Matarrese S.,  2016, \mn@doi [Riv.
  Nuovo Cim.] {10.1393/ncr/i2016-10127-1}, 39, 399

\bibitem[\protect\citeauthoryear{Howlett, Lewis, Hall  \& Challinor}{Howlett
  et~al.}{2012}]{Howlett:2012mh}
Howlett C.,  Lewis A.,  Hall A.,   Challinor A.,  2012, \mn@doi [JCAP]
  {10.1088/1475-7516/2012/04/027}, 1204, 027

\bibitem[\protect\citeauthoryear{Hu, Seljak, White  \& Zaldarriaga}{Hu
  et~al.}{1998}]{Hu:1997mn}
Hu W.,  Seljak U.,  White M.~J.,   Zaldarriaga M.,  1998, \mn@doi [Phys. Rev.]
  {10.1103/PhysRevD.57.3290}, D57, 3290

\bibitem[\protect\citeauthoryear{Huang}{Huang}{2007}]{Huang:2007qz}
Huang Q.-G.,  2007, \mn@doi [Phys. Rev.] {10.1103/PhysRevD.76.061303}, D76,
  061303

\bibitem[\protect\citeauthoryear{Huang}{Huang}{2015}]{Huang:2015xda}
Huang Q.-G.,  2015, \mn@doi [Phys. Rev.] {10.1103/PhysRevD.91.123532}, D91,
  123532

\bibitem[\protect\citeauthoryear{Huang \& Wang}{Huang \&
  Wang}{2015}]{Huang:2015gka}
Huang Q.-G.,  Wang S.,  2015, \mn@doi [JCAP] {10.1088/1475-7516/2015/06/021},
  1506, 021

\bibitem[\protect\citeauthoryear{Huang, Wang  \& Zhao}{Huang
  et~al.}{2015}]{Huang:2015gca}
Huang Q.-G.,  Wang S.,   Zhao W.,  2015, \mn@doi [JCAP]
  {10.1088/1475-7516/2015/10/035}, 1510, 035

\bibitem[\protect\citeauthoryear{Huang, Wang  \& Wang}{Huang
  et~al.}{2016}]{Huang:2015cke}
Huang Q.-G.,  Wang K.,   Wang S.,  2016, \mn@doi [Phys. Rev.]
  {10.1103/PhysRevD.93.103516}, D93, 103516

\bibitem[\protect\citeauthoryear{Kamionkowski \& Kovetz}{Kamionkowski \&
  Kovetz}{2016}]{Kamionkowski:2015yta}
Kamionkowski M.,  Kovetz E.~D.,  2016, \mn@doi [Ann. Rev. Astron. Astrophys.]
  {10.1146/annurev-astro-081915-023433}, 54, 227

\bibitem[\protect\citeauthoryear{Kamionkowski, Kosowsky  \&
  Stebbins}{Kamionkowski et~al.}{1997a}]{Kamionkowski:1996zd}
Kamionkowski M.,  Kosowsky A.,   Stebbins A.,  1997a, \mn@doi [Phys. Rev.
  Lett.] {10.1103/PhysRevLett.78.2058}, 78, 2058

\bibitem[\protect\citeauthoryear{Kamionkowski, Kosowsky  \&
  Stebbins}{Kamionkowski et~al.}{1997b}]{Kamionkowski:1996ks}
Kamionkowski M.,  Kosowsky A.,   Stebbins A.,  1997b, \mn@doi [Phys. Rev.]
  {10.1103/PhysRevD.55.7368}, D55, 7368

\bibitem[\protect\citeauthoryear{Kesden, Cooray  \& Kamionkowski}{Kesden
  et~al.}{2002}]{Kesden:2002ku}
Kesden M.,  Cooray A.,   Kamionkowski M.,  2002, \mn@doi [Phys. Rev. Lett.]
  {10.1103/PhysRevLett.89.011304}, 89, 011304

\bibitem[\protect\citeauthoryear{Khoury, Ovrut, Steinhardt  \& Turok}{Khoury
  et~al.}{2001}]{Khoury:2001wf}
Khoury J.,  Ovrut B.~A.,  Steinhardt P.~J.,   Turok N.,  2001, \mn@doi [Phys.
  Rev.] {10.1103/PhysRevD.64.123522}, D64, 123522

\bibitem[\protect\citeauthoryear{Knox}{Knox}{1995}]{Knox:1995dq}
Knox L.,  1995, \mn@doi [Phys. Rev.] {10.1103/PhysRevD.52.4307}, D52, 4307

\bibitem[\protect\citeauthoryear{Knox \& Song}{Knox \&
  Song}{2002}]{Knox:2002pe}
Knox L.,  Song Y.-S.,  2002, \mn@doi [Phys. Rev. Lett.]
  {10.1103/PhysRevLett.89.011303}, 89, 011303

\bibitem[\protect\citeauthoryear{Kobayashi, Yamaguchi  \& Yokoyama}{Kobayashi
  et~al.}{2011}]{Kobayashi:2011nu}
Kobayashi T.,  Yamaguchi M.,   Yokoyama J.,  2011, \mn@doi [Prog. Theor. Phys.]
  {10.1143/PTP.126.511}, 126, 511

\bibitem[\protect\citeauthoryear{Lasky et~al.}{Lasky
  et~al.}{2016}]{Lasky:2015lej}
Lasky P.~D.,  et~al., 2016, \mn@doi [Phys. Rev.] {10.1103/PhysRevX.6.011035},
  X6, 011035

\bibitem[\protect\citeauthoryear{Lee, Su  \& Baumann}{Lee
  et~al.}{2015}]{Lee:2014cya}
Lee H.,  Su S.~C.,   Baumann D.,  2015, \mn@doi [JCAP]
  {10.1088/1475-7516/2015/02/036}, 1502, 036

\bibitem[\protect\citeauthoryear{Lewis \& Challinor}{Lewis \&
  Challinor}{2006}]{Lewis:2006fu}
Lewis A.,  Challinor A.,  2006, \mn@doi [Phys. Rept.]
  {10.1016/j.physrep.2006.03.002}, 429, 1

\bibitem[\protect\citeauthoryear{Lewis, Challinor  \& Lasenby}{Lewis
  et~al.}{2000}]{Lewis:1999bs}
Lewis A.,  Challinor A.,   Lasenby A.,  2000, \mn@doi [Astrophys. J.]
  {10.1086/309179}, 538, 473

\bibitem[\protect\citeauthoryear{Liddle \& Lyth}{Liddle \&
  Lyth}{1992}]{Liddle:1992wi}
Liddle A.~R.,  Lyth D.~H.,  1992, \mn@doi [Phys. Lett.]
  {10.1016/0370-2693(92)91393-N}, B291, 391

\bibitem[\protect\citeauthoryear{Linde}{Linde}{1982}]{Linde:1981mu}
Linde A.~D.,  1982, \mn@doi [Phys. Lett.] {10.1016/0370-2693(82)91219-9}, 108B,
  389

\bibitem[\protect\citeauthoryear{Lyth}{Lyth}{1997}]{Lyth:1996im}
Lyth D.~H.,  1997, \mn@doi [Phys. Rev. Lett.] {10.1103/PhysRevLett.78.1861},
  78, 1861

\bibitem[\protect\citeauthoryear{Masui \& Pen}{Masui \&
  Pen}{2010}]{Masui:2010cz}
Masui K.~W.,  Pen U.-L.,  2010, \mn@doi [Phys. Rev. Lett.]
  {10.1103/PhysRevLett.105.161302}, 105, 161302

\bibitem[\protect\citeauthoryear{Meerburg, Hlo?ek, Hadzhiyska  \&
  Meyers}{Meerburg et~al.}{2015}]{Meerburg:2015zua}
Meerburg P.~D.,  Hlo?ek R.,  Hadzhiyska B.,   Meyers J.,  2015, \mn@doi [Phys.
  Rev.] {10.1103/PhysRevD.91.103505}, D91, 103505

\bibitem[\protect\citeauthoryear{Mortonson \& Seljak}{Mortonson \&
  Seljak}{2014}]{Mortonson:2014bja}
Mortonson M.~J.,  Seljak U.,  2014, \mn@doi [JCAP]
  {10.1088/1475-7516/2014/10/035}, 1410, 035

\bibitem[\protect\citeauthoryear{Mukhanov, Feldman  \& Brandenberger}{Mukhanov
  et~al.}{1992}]{Mukhanov:1990me}
Mukhanov V.~F.,  Feldman H.~A.,   Brandenberger R.~H.,  1992, \mn@doi [Phys.
  Rept.] {10.1016/0370-1573(92)90044-Z}, 215, 203

\bibitem[\protect\citeauthoryear{Page et~al.}{Page et~al.}{2007}]{Page:2006hz}
Page L.,  et~al., 2007, \mn@doi [Astrophys. J. Suppl.] {10.1086/513699}, 170,
  335

\bibitem[\protect\citeauthoryear{Renzi, Cabass, Di~Valentino, Melchiorri  \&
  Pagano}{Renzi et~al.}{2018}]{Renzi:2018dbq}
Renzi F.,  Cabass G.,  Di~Valentino E.,  Melchiorri A.,   Pagano L.,  2018,
  \mn@doi [JCAP] {10.1088/1475-7516/2018/08/038}, 1808, 038

\bibitem[\protect\citeauthoryear{Rubakov, Sazhin  \& Veryaskin}{Rubakov
  et~al.}{1982}]{Rubakov:1982df}
Rubakov V.~A.,  Sazhin M.~V.,   Veryaskin A.~V.,  1982, \mn@doi [Phys. Lett.]
  {10.1016/0370-2693(82)90641-4}, 115B, 189

\bibitem[\protect\citeauthoryear{Santos, Wang  \& Zhao}{Santos
  et~al.}{2016}]{Zhao:2015sla}
Santos L.,  Wang K.,   Zhao W.,  2016, \mn@doi [JCAP]
  {10.1088/1475-7516/2016/07/029}, 1607, 029

\bibitem[\protect\citeauthoryear{Sato}{Sato}{1981}]{Sato:1980yn}
Sato K.,  1981, Mon. Not. Roy. Astron. Soc., 195, 467

\bibitem[\protect\citeauthoryear{Seljak \& Hirata}{Seljak \&
  Hirata}{2004}]{Seljak:2003pn}
Seljak U.,  Hirata C.~M.,  2004, \mn@doi [Phys. Rev.]
  {10.1103/PhysRevD.69.043005}, D69, 043005

\bibitem[\protect\citeauthoryear{Smith, Hanson, LoVerde, Hirata  \& Zahn}{Smith
  et~al.}{2012}]{Smith:2010gu}
Smith K.~M.,  Hanson D.,  LoVerde M.,  Hirata C.~M.,   Zahn O.,  2012, \mn@doi
  [JCAP] {10.1088/1475-7516/2012/06/014}, 1206, 014

\bibitem[\protect\citeauthoryear{Starobinsky}{Starobinsky}{1979}]{Starobinsky:1979ty}
Starobinsky A.~A.,  1979, JETP Lett., 30, 682

\bibitem[\protect\citeauthoryear{Starobinsky}{Starobinsky}{1980}]{Starobinsky:1980te}
Starobinsky A.~A.,  1980, \mn@doi [Phys. Lett.] {10.1016/0370-2693(80)90670-X},
  B91, 99

\bibitem[\protect\citeauthoryear{Wang, Cai, Liu  \& Piao}{Wang
  et~al.}{2017}]{Wang:2016tbj}
Wang Y.-T.,  Cai Y.,  Liu Z.-G.,   Piao Y.-S.,  2017, \mn@doi [JCAP]
  {10.1088/1475-7516/2017/01/010}, 1701, 010

\bibitem[\protect\citeauthoryear{Zaldarriaga \& Seljak}{Zaldarriaga \&
  Seljak}{1997}]{Zaldarriaga:1996xe}
Zaldarriaga M.,  Seljak U.,  1997, \mn@doi [Phys. Rev.]
  {10.1103/PhysRevD.55.1830}, D55, 1830

\bibitem[\protect\citeauthoryear{Zhao \& Baskaran}{Zhao \&
  Baskaran}{2009}]{Zhao:2009mj}
Zhao W.,  Baskaran D.,  2009, \mn@doi [Phys. Rev.]
  {10.1103/PhysRevD.79.083003}, D79, 083003

\bibitem[\protect\citeauthoryear{Zhao \& Huang}{Zhao \&
  Huang}{2011}]{Zhao:2011zb}
Zhao W.,  Huang Q.-G.,  2011, \mn@doi [Class. Quant. Grav.]
  {10.1088/0264-9381/28/23/235003}, 28, 235003

\makeatother
\end{thebibliography}








\bsp	
\label{lastpage}
\end{document}